\RequirePackage[]{graphicx} % Loaded here to show images in draft mode, too

\def\mode{2} % 2 is Arxiv

\if 1\mode
\documentclass[draftcls,journal,web]{ieeecolor_revision}
\fi

\if 2\mode
	\documentclass[final,journal]{article}
\fi

\if 3\mode
	\documentclass[num-refs]{wiley-article_final}	
\fi

\usepackage{etoolbox}
\newtoggle{IEEE}
\newtoggle{MRM}
\newtoggle{ARXIV}

\newtoggle{SUPMATERIAL}
\togglefalse{SUPMATERIAL}

\if 1\mode
	\toggletrue{IEEE}
	\togglefalse{MRM}
	\togglefalse{ARXIV}
\fi

\if 2\mode
	\toggletrue{ARXIV}
	\togglefalse{IEEE}
	\togglefalse{MRM}
\fi

\if 3\mode
	\toggletrue{MRM}
	\togglefalse{IEEE}
	\togglefalse{ARXIV}
\fi

\iftoggle{MRM}{
	\usepackage[nomarkers, nolists]{endfloat}
}

\iftoggle{IEEE}{
}

\usepackage[utf8]{inputenc}
\usepackage{amsmath}
\usepackage{bm}
\usepackage{bbold}
\usepackage{authblk}
\usepackage{amssymb}
\usepackage{graphicx}
\usepackage{ifdraft}
\usepackage{marginnote}
\usepackage{pdfpages}
\usepackage{tabularx,booktabs}
\usepackage{ifpdf}
\ifpdf
	\usepackage[]{hyperref}
\else
	\usepackage[dvipdfm]{hyperref}
\fi
\usepackage{tikz}
\usepackage{placeins}
\usepackage[capitalise]{cleveref}
\usepackage[
per-mode=fraction,
separate-uncertainty,
retain-unity-mantissa=false,
product-units=power,
table-alignment=center,
detect-all,
binary-units=true,
exponent-product=\cdot,
exponent-base=10
]{siunitx}
\DeclareMathOperator*{\argmin}{arg\,min}

\usepackage{color}

\usepackage{changes}
\newcommand{\stkout}[1]{\ifmmode\text{\sout{\ensuremath{#1}}}\else\sout{#1}\fi}
\setdeletedmarkup{\stkout{#1}}

\title{{Cardiac and Respiratory Self-Gating in Radial MRI using an Adapted 
Singular Spectrum Analysis (SSA-FARY)}}
\newcommand{\authorA}{Sebastian Rosenzweig}
\newcommand{\authorB}{Nick Scholand}
\newcommand{\authorC}{H. Christian M. Holme}
\newcommand{\authorD}{Martin Uecker}

\newcommand{\affilA}{Institute for Diagnostic and Interventional Radiology,
	University Medical Center Göttingen, Göttingen, Germany}
\newcommand{\affilB}{German Centre for Cardiovascular Research (DZHK),
	Partner Site Göttingen, Göttingen, Germany}

\newcommand{\corAdress}{Sebastian Rosenzweig, University Medical Center 
Göttingen, Institute for Diagnostic and Interventional Radiology, 
Robert-Koch-Str. 40, 37075 Göttingen, Germany. }

\newcommand{\corMail}{sebastian.rosenzweig@med.uni-goettingen.de}
\newcommand{\runHead}{Rosenzweig et al.}
\newcommand{\Funding}{
	Supported by the DZHK (German Centre for Cardiovascular 
	Research), the Physics-to-Medicine Initiative Göttingen (LM der 
	Niedersächsischen Vorab), the DFG (German Research Foundation) under
	grant UE 189/1-1, and the Studienstiftung des deutschen Volkes.
	We gratefully acknowledge the support of the
	NVIDIA Corporation with the donation of one
	NVIDIA TITAN Xp GPU for this research.
}

\nottoggle{IEEE}{
		
		\iftoggle{ARXIV}{
				\author[1,2]{\authorA \thanks{\corAdress \corMail}}
			}
		
		\iftoggle{MRM}{
				\author[1,2]{\authorA}
			}
		
		\author[1,2]{\authorB}
		\author[1,2]{\authorC}
		\author[1,2]{\authorD}
		
		\affil[1]{\affilA}
		\affil[2]{\affilB}

		\iftoggle{MRM}{
				\papertype{Note}
				\corraddress{\corAdress}
				\corremail{\corMail\\
						
						Submitted to XXX
						\bf{{Word count}}: Abstract , Body .}
				
				\runningauthor{\runHead}
			}
}

\iftoggle{IEEE}{
	\usepackage{tmi}
	\usepackage{cite}
	\usepackage{amsmath,amssymb,amsfonts}
	\usepackage{algorithmic}
	\usepackage{graphicx}
	\usepackage{textcomp}
	\author{
		{\authorA, \authorB, \authorC, \authorD}		
		\thanks{Date submitted for review: 2020-02-18.}
		\thanks{\Funding}		
		\thanks{\authorA: \affilA, \affilB. Corresponding author \corMail.}
		\thanks{\authorB: \affilA, \affilB.}		
		\thanks{\authorC: \affilA, \affilB.}	
		\thanks{\authorD: \affilA, \affilB.}
					
	} 	
}

\begin{document}
	
\maketitle

\begin{abstract}
Cardiac Magnetic Resonance Imaging (MRI) is
time-consuming and error-prone. 
To ease the patient's burden and to increase the efficiency
and robustness of cardiac exams, interest in methods based on
continuous steady-state acquisition and self-gating has been
growing in recent years.
Self-gating methods extract the cardiac and respiratory signals from
the measurement data and then retrospectively sort the data
into cardiac and respiratory phases. Repeated breathholds and
synchronization with the heartbeat using some external device
as required in conventional MRI are then not necessary. In this work, we introduce
a novel self-gating method for radially acquired data
based on a dimensionality reduction technique for
time-series analysis (SSA-FARY). Building on Singular Spectrum Analysis,
a zero-padded, time-delayed embedding of the auto-calibration 
data is analyzed
using Principle Component Analysis.
We demonstrate the basic functionality of SSA-FARY using numerical
simulations and apply it to in vivo cardiac radial single-slice bSSFP and 
Simultaneous
Multi-Slice radio-frequency-spoiled gradient-echo measurements, as well as to
Stack-of-Stars bSSFP measurements.
SSA-FARY reliably detects the cardiac and respiratory motion and separates
it from noise. We utilize the generated 
signals for
high-dimensional image reconstruction using parallel imaging and compressed sensing
with in-plane wavelet and (spatio-)temporal total-variation
regularization.

\end{abstract}

\section{Introduction}

Magnetic Resonance Imaging (MRI) is an intrinsically slow 
imaging technique,  
which makes imaging of moving organs particularly challenging.  Still, from the early years of 
MRI, researchers recognized the great chances and implications of monitoring 
the beating heart without the use of ionizing radiation and
with the superior tissue contrast of MRI. 
Here, the respiratory and cardiac motion pose high demands
on the acquisition and reconstruction.

One solution is real-time imaging, which resolves the true 
dynamics of the 
heart but is limited in terms of temporal and spatial resolution and
restricted to two-dimensional imaging \cite{Kerr_Magn.Reson.Med._1997,  Yang_JAmCollCardiol_1998, Uecker_Magn.Reson.Med._2010,Uecker_NMRBiomed._2010}. %Poddar_IEEETrans.Med.Imag._2016
In clinical practice, pro- or retrospective gating is typically used, which exploits the 
quasi-periodicity of the respiratory and cardiac motion to compose a single 
synthetic heartbeat from data acquired during several actual beats.
To synchronize data-acquisition with 
breathing motion, external devices like a
respiratory belt or adapted sequences with navigator readouts are commonly used 
\cite{Ehman_AmJRoentgenol_1984, Liu_Magn.Reson.Med._1993}. However, these 
devices have to be placed and adjusted 
individually for each patient. Furthermore, the resulting respiratory signal is
not always 
directly correlated to the motion of the heart \cite{Guo__2017}.
Sequences with additional interleaved navigator acquisitions prolong the 
measurement substantially and complicate the use of steady-state sequences.
Otherwise, breath-hold commands can be used to avoid the need for respiratory 
gating completely, which, however, can be exhausting, time-consuming and not
expedient for sick or non-compliant patients and children.
For cardiac gating, the standard in clinical practice is the use of 
an electrocardiogram (ECG) \cite{Higgins__1987}, but the ECG signals
can experience signal distortion when MRI sequences with fast
gradient switching are utilized \cite{Rokey_MagnResonMed_1988}.

To avoid these drawbacks and to gain more flexibility, techniques have been
developed to extract cardiac motion from the 
data itself, which is known as retrospective \textit{self-gating} 
\cite{Larson_Magn.Reson.Med._2004}. Similar approaches can also be used to 
extract respiratory motion \cite{Kim_MagnResonMed_1990}. A large 
number of 
different strategies for cardiac, respiratory or combined self-gating with 
Cartesian or Non-Cartesian acquisition were proposed in the past, e.g. 
\cite{Larson_Magn.Reson.Med._2004, Crowe_Magn.Reson.Med._2004, 
Larson_Magn.Reson.Med._2005, Uribe_Magn.Reson.Med._2007, 
Buehrer_Magn.Reson.Med._2008, 
Paul_Magn.Reson.Med._2015}. Still, the fundamental idea in most approaches is 
similar: Either a 1D signal is extracted from certain receive channels using
a band-pass filter and specific properties of the acquired auto-calibration (AC)
data, or a (sliding window) low-spatial 
high-temporal resolution reconstruction of a specific region of interest (ROI) 
is analyzed. A more sophisticated yet simple idea was proposed by Pang et al. 
\cite{Pang_Magn.Reson.Med._2014}: The general concept of dimensionality 
reduction \cite{Kirby__2000} is applied to the AC data by using a
Principle Component Analysis (PCA) to extract the required motion signals.
However, the resulting signals are often spoiled by noise or 
trajectory-dependent oscillations, which makes additional filtering necessary 
\cite{Deng_Magn.Reson.Med._2016, Feng_Magn.Reson.Med._2016a, Rosenzweig_Proceedingsofthe26thAnnualMeetingofISMRM_2018}. Moreover, cardiac 
and respiratory motion are not always clearly separated 
\cite{Gao_J.Cardiov.Magn.Reson._2016}, which complicates data binning into the 
respective breathing and heart phases and requires the use of further 
post-processing steps such as coil clustering \cite{Zhang_Magn.Reson.Med._2016}.

To overcome these limitations, we propose the use of an adapted Singular 
Spectrum Analysis (SSA), which can be thought of as a temporally localized PCA 
or equivalently as a PCA applied to time-delay embedded 
coordinates.
SSA is an application of the 
general Karhunen-Loève theorem \cite{Fukunaga__2013} and a powerful tool for 
the analysis of dynamical 
systems, incorporating elements of classical time-series analysis, multivariate 
statistics,
multivariate geometry and signal processing 
\cite{Golyandina__2001}. Broomhead and King 
 derived SSA from Takens' theorem for the 
analysis of chaotic dynamical systems, and applied it to the 
problems of dynamical systems theory \cite{Takens__1981, Broomhead_PhysicaD_1986}. Further development was promoted by 
Vautard et al. \cite{Vautard_PhysicaD_1989, 
Vautard_PhysicaD_1992}. Since the birth of SSA in 1986 
\cite{Broomhead_PhysicaD_1986, Broomhead__1986} it has found 
wide-spread application in various fields \cite{Hannachi_Int.J.Climatol._2007, 
Koelle_Nature_2005,Rezek_IEEETrans.Biomed.Eng._1998, 
Kumar_Energy_2010, Wu_J.Hydrol._2009, Janssens_NatGenet_2006}.
SSA can be used for noise 
reduction, detrending and the identification of 
oscillatory components \cite{Vautard_PhysicaD_1992}, hence it is ideally suited 
for the extraction of vital motion signals such as respiratory and cardiac 
motion in self-gated MRI. 

Nevertheless, conventional univariate SSA can only be applied 
to single-channel time series, whereas in parallel MRI multiple receive 
channels (phased array coils) are available. Channels located closer to 
the heart tend to capture cardiac motion, while coils placed near the diaphragm
rather monitor respiratory motion. Manual coil selection 
\cite{Feng_J.Cardiov.Magn.Reson._2014} can enable the use of univariate SSA 
for vital motion extraction, but correlated information from other coils 
is then lost. Moreover, for routine clinical use a 
fully automated technique is preferred.

Fortunately, univariate SSA has a natural extension for the 
analysis of a multi-channel time series \cite{Broomhead__1986}. However, this
multivariate SSA is not a 
dimensionality reduction technique, but recovers the specific oscillations for 
each channel rather than to extract a single signal that describes  
the temporal evolution of the principle motion components. 

Here, we adapt the Singular Spectrum Analysis For Advanced Reduction of 
dimensionalitY, which we dub 
SSA-FARY \cite{Rosenzweig__2019a, Rosenzweig__2019b}. In its original 
form (multivariate) SSA 
consists of four steps 
\cite{Golyandina__2001}: I) Hankelization, II) Decomposition, III) Grouping, 
IV) Backprojection. In SSA-FARY, we remove steps III) and IV) and 
instead
perform a  zero-padding operation at the start. 
We will demonstrate the basic functionality of SSA-FARY in numerical 
simulations and show reconstructions of in vivo cardiac measurements
acquired with single-slice bSSFP and Simultaneous Multi-Slice (SMS) 
radio-frequency (RF)-spoiled gradient-echo (FLASH)
sequences, as well as with a Stack-of-Stars (SOS) bSSFP sequence.

\section{Theory}

In radial single-slice, SMS or Stack-of-Stars imaging the central k-space point
or the central line along the slice-dimension $k_z$ ($k_x=0$, $k_y=0$), respectively, 
have proved to be ideally suited for self-gating 
\cite{Larson_Magn.Reson.Med._2004, Feng_Magn.Reson.Med._2016a}. 
We extract this AC data from the measurement data and stack all coils and 
partitions into a single dimension. This yields a multi-channel time-series 
${X}_{c}^{\;t}$ of size $[(N_p \times N_c) \times N_{t}]$, with $1\leq t \leq 
N_{t}$ and $1 \leq c \leq (N_p\cdot N_c)$, which contains information about
the respiratory and cardiac motion. $N_t$ is the total number of central 
k-space points or lines used for auto-calibration, $N_p$ is the number of 
partitions and $N_c$ is the number of receive coils. Each channel $c$ is 
normalized 
to have zero mean.

\subsection{Correction of the AC Data}

System imperfections such as gradient delays and off-resonances usually cause a 
corruption of the AC data $\bm{X}$, which manifests an oscillation of a
trajectory-dependent frequency in radial imaging \cite{Deng_Magn.Reson.Med._2016}.
This  signal fluctuation is often misinterpreted by dimensionality reduction methods 
as a major signal contribution. This contribution can mostly be removed by 
a simple orthogonal projection based on its known frequency.
Here, we extend this method to also include higher-order harmonics
which yields a method that almost completely removes the unwanted signal.
For simplicity, we assume a golden angle acquisition scheme.
Let $\varphi_0$ be the incremental projection angle, then 
\begin{equation}
\varphi^t = t \cdot \varphi_0
\label{Eq:Phi}
\end{equation} 
is the projection angle used for the acquisition at time step $t$. 
We define the vector
\begin{equation}
	n^{t} := \left( \begin{array}{c}
	{e^{i\varphi^t}} \\
	{e^{-i\varphi^t}} \\
	{e^{i\cdot 2\varphi^t}} \\
	{e^{-i\cdot 2\varphi^t}} \\
	\vdots \\
	{e^{i\cdot N_H\varphi^t}} \\
	{e^{-i\cdot N_H\varphi^t}} \\	
	\end{array}\right),
\end{equation}
containing the oscillations up to the $N_H$-th harmonic 
as a basis for the perturbing oscillation and constrain 
$\bm{X}$ to be orthogonal to $\bm{n}$,
\begin{equation}
\bm{X}_\text{cor} = \bm{X}_\text{raw} - \bm{n}(\bm{n}^\dagger 
\bm{X}_\text{raw}),
\end{equation}
with $^\dagger$ denoting the pseudo-inverse. This procedure cleans the 
corrupted signal $\bm{X}_\text{raw}$ and yields a
corrected time series $\bm{X}_\text{cor}$. We use this AC correction
method in all presented in vivo experiments.

\subsection{Dimensionality Reduction Methods}

\paragraph{Principle Component Analysis}

PCA can be understood as the rotation of the original coordinate system to a 
new one with orthogonal axes that coincide with the directions of 
maximum variable variance \cite{Campbell_SystematicZoology_1981}. The PCA of a 
time series $\bm{X}$ can be performed using the Singular Value Decomposition 
(SVD).
\begin{equation}
\bm{X}^T = \bm{U}\bm{S} \bm{V}^H.
\end{equation}
Here, the diagonal matrix $\bm{S}$ contains the real eigenvalues 
$ \lambda_1 \geq \dots \geq \lambda_{{N}_{t}} \geq 0$ in decreasing 
order of  magnitude.
PCA provides the expansion of $\bm{X}^T$ onto the 
orthonormal [$N_{t} \times N_{t}$] basis $\bm{U}$,
\begin{equation}
(X^T)_{t} = \sum_{k=1}^{N_{t}} {U}_t^{\;k}(SV^H)_{k},
\end{equation} 
where the \textit{principle components} $(SV^H)_{k}$ are given by 
\begin{equation}
(SV^H)_{k} := \sum_{l=1} ^{\; N_c \cdot N_p} S_{k}^{\;l} (V^H)_l = 
\lambda_k (V^H)_k~.
\end{equation}
Since the cardiac and respiratory motion signals contribute as main sources of 
variation to the time series
$\bm{X}$, their temporal behavior should be 
captured by one of the first basis vectors $U^{k}$, respectively 
\cite{Pang_Magn.Reson.Med._2014}.

%NOTE: I remove the semicolon notion and the c index. --> OK

\paragraph{Singular Spectrum Analysis For Advanced Dimensionality 
Reduction (SSA-FARY)}

A schematic of the SSA-FARY procedure is depicted in Fig.\ \ref{Fig:SSA}a. 
In contrast to conventional (multivariate) SSA, we first zero-pad
($\mathcal{Z}$) the second dimension of the AC data 
$\bm{X}$ to obtain matrix $\tilde{\bm{X}}$ of size 
[$(N_p \times N_c) \times ({N}_{t} + W - 1) $],
\begin{equation}
\tilde{\bm{X}} = \mathcal{Z}{\bm{X}}.
\label{Eq:ZeroPad}
\end{equation}
Next, we construct a Block-Hankel calibration matrix 
\begin{equation}
\bm{A}= \mathcal{H}\tilde{\bm{X}}
\label{Eq:Hankel}
\end{equation} 
of size 
[$ N_{t} \times ((N_p \times N_c)\times W)$]. Here, 
the Hankelization operator $\mathcal{H}$ slides a window of size [$1 \times W$]
through channel $\tilde{X}_c$ of the zero-padded AC data and 
takes each block to 
be a row in the $c$-th column of the calibration matrix. This operation is 
similar to the construction of the calibration matrix in ESPIRiT 
\cite{Uecker_Magn.Reson.Med._2014}.
	
We decompose $\bm{A}$ using an SVD
\begin{equation}
\bm{A} = \bm{U}\bm{S} \bm{V}^H,
\end{equation}
and consider $\bm{U}$ of size [${N}_{t} \times {N}_{t}$] as the 
orthonormal basis that consists of the 
\textit{Empirical Orthonormal Functions} (EOFs) $U^{k}$, $1 \leq 
k \leq {N}_{t}$. 
The \textit{principle components} $(SV^H)_{k}$ are given by
\begin{equation}
(SV^H)_{k} := \sum_{l=1} ^{\; W \cdot N_c \cdot N_p}	S_{k}^{\;l} 
(V^H)_l = \lambda_k (V^H)_k.
\end{equation} 
The expansion of $\bm{A}$, or $\tilde{\bm{X}}$, in the basis 
$\bm{U}$ then reads
\begin{equation}
\tilde{X}_{c}^{\; {t}+j} = A_{{t}}^{\; cj}= \sum_{k=1}^{{N}_{t}} 
U_{{t}}^{\; k} (SV^H)_{k}^{\; cj},
\end{equation}
where $1 \leq {t} \leq {N}_{t}$ iterates through the
temporal samples, $1 \leq c \leq (N_c\cdot N_p)$ through the
channels and $0 \leq j < W$ is the index inside the sliding window. 
The EOFs can be considered as data-adaptive weighted moving 
averages of the original time series $\tilde{\bm{X}}$, with $\bm{V}$
being the data-adaptive filters \cite{Vautard_PhysicaD_1992, 
Harris_PhysicaD_2010},
\begin{equation}
\bm{U}\bm{S} = \bm{A}\bm{V},
\label{Eq:MovingAvg}
\end{equation}
\begin{equation}
U_t^{\; k} = \frac{1}{\lambda_k}\sum_{c=1}^{N_p \cdot N_c} \sum_{j=1}^W 
\tilde{X}_c^{\; t+j} V_{cj}^{\; k}. 
\label{Eq: Eigenfilter}
\end{equation}
In fact, the columns of $\bm{V}$ can bee seen as a complete eigenfilter 
decomposition of the original time series \cite{Kume_Adv.Adapt.DataAnal._2012}. 
These filters $V^k$ act as data-adaptive band-pass filters with a 
frequency bandwidth $\delta f_\text{B}$ given by
\begin{equation}
\delta f_\text{B} = \frac{f_\text{s}}{W},
\label{Eq:BP}
\end{equation}
where $f_\text{s}$ is the sampling rate \cite{Leles__2016, 
	Xu_Sensors_2018}. Harris and Yuan showed in 
\cite{Harris_PhysicaD_2010} for the univariate case that a periodic 
oscillation contained in the data lead to an even and odd filter. The 
application of these filters to the original time series constitutes for each 
oscillation one EOF 
which is in phase and one which is in quadrature to the original oscillation, 
respectively.

Vautard et al. \cite{Vautard_PhysicaD_1992} proposed another interpretation 
of the EOFs considering the minimization problem
\begin{equation}
\argmin_{\alpha}|_{t,k} \sum_{c=1}^{N_p \cdot N_c} \sum_{j=1}^W \left\lVert 
\tilde{X}_c^{\; 
t+j}- 
\alpha 
(SV^H)_k^{\; cj} \right\rVert^2.
\label{Eq:Minimization}
\end{equation}
The solution of eq.\ (\ref{Eq:Minimization}) is $\alpha=U_t^{\; k}$, 
thus the EOFs can be obtained by a local least-squares fit of the k-th 
principle component to the original time series. This locality, determined
by the window size $W$, distinguishes SSA-FARY from classical PCA, which 
does not take the temporal past and future of a sample into account. In 
fact, PCA is a special case of SSA-FARY with $W=1$,	
\begin{equation}
\bm{A}|_{W=1} = \mathcal{H}{\tilde{\bm{X}}} = 
\mathcal{H}\mathcal{Z}\bm{X} = \bm{X}^T.
\end{equation}

In conventional SSA, i.e. where no zero-padding is applied, the EOFs are of 
reduced length $N_t - W + 1$. Thus, the exact correspondence in time is 
lost, which inhibits their use as self-gating signal. In contrast,
the EOFs $U^{k}$ in SSA-FARY preserve the length $N_t$ of the 
original time series and can directly be used for self-gating, similar to 
the eigenvectors $U^{k}$ in PCA. In distinction to PCA, the EOFs in SSA(-FARY)
capture  temporal oscillations via oscillatory pairs \cite{Vautard_PhysicaD_1989}. 
In particular, if two consecutive eigenvalues are nearly equal, the two 
corresponding EOFs are nearly periodic with the same period and in 
quadrature \cite{Plaut_JAtmosSci_1994}, which is a consequence of the 
filtering 
property of SSA-FARY \cite{Harris_PhysicaD_2010}. To ensure a proper separation 
the singular values of different EOF pairs should be distinct, which is called 
the strong separability condition \cite{Golyandina_Stat.Interface_2015} and 
usually fulfilled for our application.

These pairs can be seen as the 
data-adaptive equivalent to the sine-cosine pairs of Fourier analysis  
\cite{Groth_Phys.Rev.E_2011}. A single EOF-pair might suffice for the
analysis of nonlinear and inharmonic oscillations, as it automatically 
locates intermittent oscillatory regions. In contrast, classical spectral 
analysis would require a large amount of harmonics or subharmonics of the 
fundamental period \cite{Vautard_PhysicaD_1992, Ghil_Rev.Geophys._2002}.
	
\paragraph{Comments}
The EOF $U^{k}$ is also the k-th left 
eigenvector of the [${N}_{t} 
\times {N}_{t}$] real-symmetric cross-covariance
matrix 
\begin{equation}
\bm{C} = \bm{A}\bm{A}^H.
\label{Eq:Cov}
\end{equation} 
Depending on the number of acquired spokes and partitions, 
computing the eigen-decomposition of 
$\bm{C}$ is usually more efficient than computing
the SVD of $\bm{A}$.

In contrast to the parameter-free PCA, for SSA we must define a window 
size $W$. To long-range correlations in time, $W$ should be large, 
which - as a trade-off - results in a lower degree of statistical 
confidence \cite{Ghil_Rev.Geophys._2002}. Vautard et al. \cite{Vautard_PhysicaD_1992} 
showed that SSA can resolve oscillations best when the periods 
are shorter than the window size $W$. In our study, the window size 
$W\approx\SI{3}{\second}$ proofed to be a robust choice for most measurements, 
independently of the utilized sequence, and this value was chosen as the 
default. More information on the choice 
of the window size is given in the Methods and Discussion section.

The fundamental concept behind the use of a temporal window $W$ is Taken's delay
embedding theorem \cite{Takens__1981}, one of the backbones of chaotic
dynamical system analysis. Instead of considering each temporal sample
individually and isolated from other time points, so called time-delay
coordinates are constructed by embedding the samples in a higher-dimensional
space with embedding dimension $W$. Consequently, each time point is
represented by a time-delay coordinate vector, which comprises not only the
sample of the respective time but also its temporal past and future. It is
therefore a natural choice to pick an odd value for $W$ in order to incorporate
the same amount of past and future information.

Towards the beginning and the end of a time series this embedding can no
longer be constructed due to lack of past or future samples, respectively.
There are two strategies to overcome this limitation: 1.) Time-delay
coordinates are constructed for the central $N_t - (W - 1)$ samples only, which
means that $W-1$ samples would be discarded from future processing. 2.) $(W -
1) / 2$  samples are zero-padded on both ends of the time series, which comes
at the expense of increased inaccuracy for the marginal samples of the
time series. However, for the second approach no samples have to be discarded
and through the symmetric zero-padding the time-delay coordinates remain in
sync with the actual temporal evolution of the signal. In this manuscript, the
second approach is used.

\subsection{Binning}	
The myocardium shows different behavior for contraction and expansion, so 
usually the entire cardiac cycle is divided into multiple distinct bins to 
accurately resolve the temporal motion. For respiratory gating it is usually
assumed that inspiration and expiration do 
not have to be distinguished 	
\cite{Liu_Magn.Reson.Med._2010,Feng_Magn.Reson.Med._2016a, 
	Paul_Magn.Reson.Med._2015}. However, various studies reveal that 
respiratory motion is 
heavily subject-dependent and exhibits a strong 
variability as well as hysteresis, which affects the global position of the 
myocardium \cite{Nehrke_Radiology_2001, 
	Burger_MagnResonMed_2013, Dasari_IEEETransNuclSci_2014, 
	Dasari_MedPhys_2017}. Hence, inspiration and expiration should be 
distinguished to properly resolve the effects of breathing motion on the 
heart.

\begin{figure}[h]
	\centering
	\includegraphics[width=0.9\textwidth]{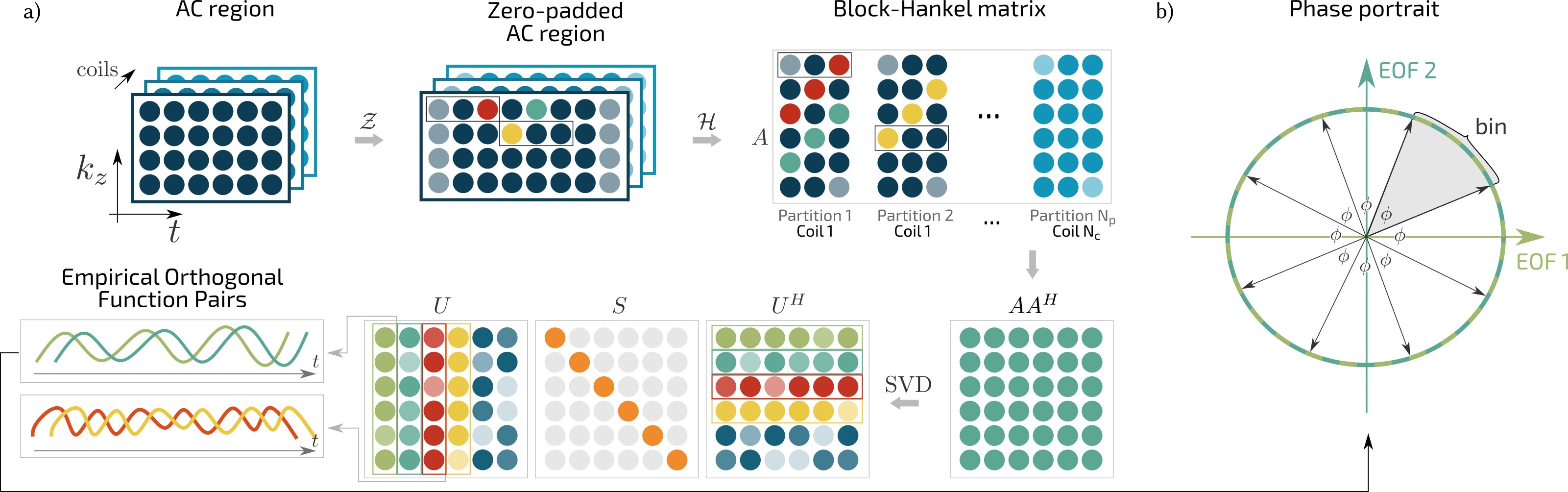}
	\caption{(a) Schematic of SSA-FARY: The multi-channel AC data is 
	zero-padded 
	$\mathcal{Z}$ and Hankelized $\mathcal{H}$. Then the cross-correlation 
	matrix $AA^H$ is calculated and decomposed using an eigenvalue or 
	singular 
	value decomposition. The left eigenvector $U$ contains the EOF quadrature 
	pairs that represent the principle motion signals of the time series. (b) 
	Binning: The phase portrait of an EOF quadrature pair is divided into $N$ 
	circular sectors with equal central angle $\varphi = 360^\circ / N$. The 
	samples are then assigned to bins according to their respective circular 
	sector.}
	\label{Fig:SSA}
\end{figure}

Since SSA-FARY yields EOF quadrature pairs that capture the phase 
information of periodic oscillations, binning is straight-forward for both 
cardiac and respiratory motion: The phase portrait, i.e.\ the 
amplitude-amplitude scatter plot, of an EOF pair is divided into $N$ 
circular sectors with central angle $\phi = \SI{360}{\degree}/{N}$. The 
samples are then binned according to their respective circular sector, see 
Fig.\ \ref{Fig:SSA}b.

\section{Methods}
\subsection{Numerical Simulations}
We compare the capability of PCA and SSA-FARY in extracting and separating 
oscillatory signals in simple numerical simulations.

The signals we want to extract are two frequency-modulated sinusoids
\begin{equation}
a(t) = A \sin(\varphi_a + \omega_a t + \phi_a(t)),
\end{equation}
\begin{equation}
b(t) = B \sin(\varphi_b + \omega_b t + \phi_b(t)),
\end{equation}
with
\begin{equation}
\phi_a(t) =  \Phi \sin(2\pi t/2T),
\end{equation}
\begin{equation}
\phi_b(t) =  \Phi \sin(2\pi t/T),
\end{equation}
which account for frequency variations.
To simulate various channels $i$, we use a weighted sum
\begin{equation}
x_i (t) = \frac{i}{N_c} a(t) + \frac{(N_c + 1 - i)}{N_c} b(t),\;\; 
1\leq i \leq N_c. 
\end{equation}
To spoil the composite signals $x_i(t)$, we add Gaussian white noise  
$h_\text{noise}(t)$
with standard deviation $\sigma_\text{noise}$,
or an oscillatory spell from time $t_1$ to time $t_2$,
\begin{equation}
h_\text{spell}(t) = C \sin(\varphi_c + \omega_c t),\;\;  
t_1 \leq t \leq t_2,
\end{equation}
or an exponential trend
\begin{equation}
h_\text{trend} (t) = D_1 e^{\xi t} - D_2,
\end{equation}
to all channels, which yields
\begin{equation}
X_i(t) = x_i(t) + h(t).
\end{equation}
We analyze the time-series $\bm{X}^\text{noise}$, $\bm{X}^\text{spell}$ and 
$\bm{X}^\text{trend}$ using PCA and SSA-FARY with window size 
$W$. 
Note, that the aim of this numerical experiment is to demonstrate the 
general benefits of SSA-FARY over PCA for the analysis of time series, and not 
to simulate cardiac and respiratory 
motion in the most accurate way. We therefore 
did not include the modeling of a more complex frequency variability or motion 
signal shapes. More details on 
the simulation are provided in the appendix.

\subsection{In Vivo experiments}
All measurements were performed on a Skyra 3T scanner (Siemens Healthcare GmbH, 
Erlangen, Germany)
using 30 channels 
of a thorax and spine coil. Gradient delay correction was performed using 
RING \cite{Rosenzweig_Magn.Reson.Med._2018a, Rosenzweig__2019}. The AC data 
was corrected using 
the orthogonal projection with $N_H=5$. In the following, the AC data's 
real and 
imaginary part are 
treated as individual channels. The field 
of 
view in all 
experiments was 
$256\times\SI{256}{\milli\meter}^2$ at base resolution $192$. All presented
experiments were performed on volunteers with no known diseases, who 
gave written informed consent. All SOS measurements were performed on 
different volunteers. The study had received approval from the local
ethics committee.

\paragraph*{Sequence Design, Auto-Calibration and Reconstruction}
We utilize a radial bSSFP sequence for the single-slice measurement, an 
RF-spoiled gradient-echo sequence with randomized RF
spoiling 
\cite{Roeloffs_Magn.Reson.Med._2016} for the SMS measurement
and a radial bSSFP sequence with undersampling in $k_z$ direction 
\cite{Feng__2016} for the SOS measurement. To obtain maximum k-space 
coverage
and thus improved image quality 
\cite{Zhou_Magn.Reson.Med._2017, Rosenzweig_Magn.Reson.Med._2018}, the 
projection angle $\varphi$ is increased \textit{in each shot} about
the seventh tiny golden angle $\varphi_0 \approx \SI{23.6}{\degree}$ 
\cite{Wundrak_IEEETransMedImag_2015}.
For the SMS and SOS measurements, the partitions are acquired in an interleaved 
fashion, i.e. one spoke is recorded for each partition before the next in-plane 
spoke of a partition is acquired. 

For single-slice imaging, the central sample 
of all spokes is used for auto-calibration. 
For SMS, more AC data is available as not only a single sample 
but the central line 
along the $k_z$ ($k_x=0,\; k_y=0$) direction can be utilized. 

Inspired by \cite{Feng__2016}, 
we make use 
of variable-density $k_z$-undersampling for the SOS acquisition. From the total 
number of 14 partitions, the central 6 partitions are always 
acquired and the corresponding central line is used for 
auto-calibration. The remaining 8 partitions are undersampled by a factor of 4. 
This center-dense sampling scheme does not only increase the temporal 
resolution of the AC data but also improves the image quality 
\cite{Berman__2016}.

For self-gating with PCA and SSA-FARY as well 
as for imaging reconstruction we use \texttt{BART} \cite{Uecker__2015}. 
Image reconstruction is performed using combined parallel
imaging and compressed sensing (PICS) \cite{Block_Magn.Reson.Med._2007} 
applying the alternating direction method
of multipliers (ADMM) \cite{Boyd_Found.TrendsMach.Learn._2011}
with in-plane wavelet-regularization on the spatial 
dimensions and total variation (TV) on the cardiac and respiratory dimension
\cite{Feng_Magn.Reson.Med._2016a,Cheng_Sci.Rep._2017}.
For SOS imaging, we additionally apply TV regularization in slice direction. 
The coil sensitivities for the single-slice and SMS measurements are generated using 
radial ENLIVE allowing two maps \cite{Uecker_Magn.Reson.Med._2010, 
	Rosenzweig_Magn.Reson.Med._2018, Holme__2017, Uecker_Magn.Reson.Med._2014}.
To reduce the memory demand we allow only one map in the SOS reconstruction. We 
apply coil compression 
\cite{Huang_Magn.Reson.Imaging_2008,Buehrer_Magn.Reson.Med._2007} to reduce the 
number of coils to 13 for 
single-slice and SMS, and 10 for SOS imaging and perform the calibration of the 
sensitivities using a lower resolution.

The SSA-FARY 
gating signals is distributed into 25 cardiac and 9 respiratory 
bins for image reconstruction. Although not 
always necessary, we standardly perform an additional 
detrending of the EOFs using a moving average filter of length $L_\text{avg} 
\approx 3 
W$. This 
further improves the 
binning accuracy of SSA-FARY by removing a possibly remaining residual trend. 

In the spirit of reproducible research, code and data to reproduce the 
experiments are made available on 
Github.\footnote{\url{https://github.com/mrirecon/SSA-FARY}}

\paragraph*{Single-slice imaging}

We perform a $72$ second free-breathing bSSFP scan (TE/TR = 
$1.90/\SI{3.80}{\milli\second}$, 
flip angle $\SI{40}{\degree}$)
with slice-thickness $\SI{7}{\milli\meter}$ 
of the human heart in short-axis view and use the first 30 seconds of data for 
further analysis and image reconstruction. The full $72$ 
second scan 
is 
used for a gridding reconstruction in Supplementary Material chapter II. We furthermore conduct an ECG-triggered CINE bSSFP breath-hold scan of the 
same slice (TE/TR =  
$1.52/\SI{3.04}{\milli\second}$, 
flip angle $\SI{57}{\degree}$, slice-thickness $\SI{7}{\milli\meter}$).

We compare the principal motion signals using PCA and 
SSA-FARY with window size $W=751$, which 
covers a 
period of about 
$\SI{3}{\second}$.

\paragraph*{SMS imaging}

We perform a 60 second free-breathing SMS RF-spoiled gradient-echo scan (TE/TR = $1.79/\SI{2.90}{\milli\second}$, 
flip angle $\SI{12}{\degree}$) 
with  three simultaneously acquired slices in short-axis view. The slice thickness 
is $\SI{5}{\milli\meter}$ and the slice gap $\SI{10}{\milli\meter}$. We use 
SSA-FARY with window size $W=345$, which 
covers a period of about $\SI{3}{\second}$. We 
perform a joint reconstruction of all slices using binning based on SSA-FARY. 
To evaluate the accuracy of SSA-FARY, we compare 
the SSA-FARY respiration quadrature-signals
of the complete time series
with the breathing pattern extracted from a pneumatic respiratory 
belt (Siemens Healthcare GmbH, Erlangen, Germany) and
a real-time reconstruction
\cite{Rosenzweig_Magn.Reson.Med._2018, Rosenzweig__2017}.

\paragraph*{SOS imaging}
We measure 8 volunteers and on each we perform a free-breathing three-minute radial SOS 
bSSFP scan (TE/TR = 
$1.90/\SI{3.80}{\milli\second}$, flip angle $\SI{35}{\degree}$) with 
fourteen partitions in short-axis view and slice thickness 
$\SI{7}{\milli\meter}$.
By default we use 
SSA-FARY with window size $W=91$, which 
covers a period of about $\SI{2.8}{\second}$. For one volunteer the 
respiratory EOF pair revealed highly irregular breathing with occasional 
breath-holds of up to $\SI{12}{\second}$ and for 
another volunteer the cardiac 
EOF pair showed a pronounced frequency variation. To improve the gating 
accuracy in these two cases, we determined another respiratory EOF pair using 
window size $W=51$
and another cardiac EOF pair using window size $W=21$, respectively. 

To evaluate the precision of the cardiac gating 
with 
SSA-FARY, we analyze the SSA-FARY signal in comparison to
the simultaneously acquired ECG 
trigger using a \texttt{Python 3} script. Therefore, we define a synthetic 
trigger point when the phase related to the orthogonal cardiac SSA-FARY 
quadrature pair 
experiences a zero-phase crossing. Since the global phase-offset of the 
quadrature pair is arbitrary, we correct the synthetic SSA-FARY trigger by a 
constant shift using the 
average distance to the ECG trigger. We then compute the standard deviation 
$\sigma_\text{trig}$ and standard error of the corrected SSA-FARY trigger to 
the ECG trigger.

Moreover, we acquire the same slices using a conventional ECG-triggered 
breath-hold CINE single-slice stack bSSFP measurement with Cartesian read-out 
(TE/TR = 
$1.52/\SI{3.04}{\milli\second}$, flip angles depending on the Specific 
Absorbtion Rate (SAR) limits between 
$\SI{47}{\degree}$ and $\SI{62}{\degree}$) and cardiac 
bin size $T_\text{bin}\approx\SI{48}{\milli\second}$. The patient dependent 
measurement time is around $6-\SI{8}{\minute}$. We compare 
the end-diastolic and end-systolic left-ventricular blood-pool area of a 
mid-ventricular slice 
to the SSA-FARY based reconstruction using \texttt{ImageJ}.

\section{Results}

\paragraph*{Numerical simulations}

\begin{figure}[h]
	\centering
	\includegraphics[width=0.40\textwidth]{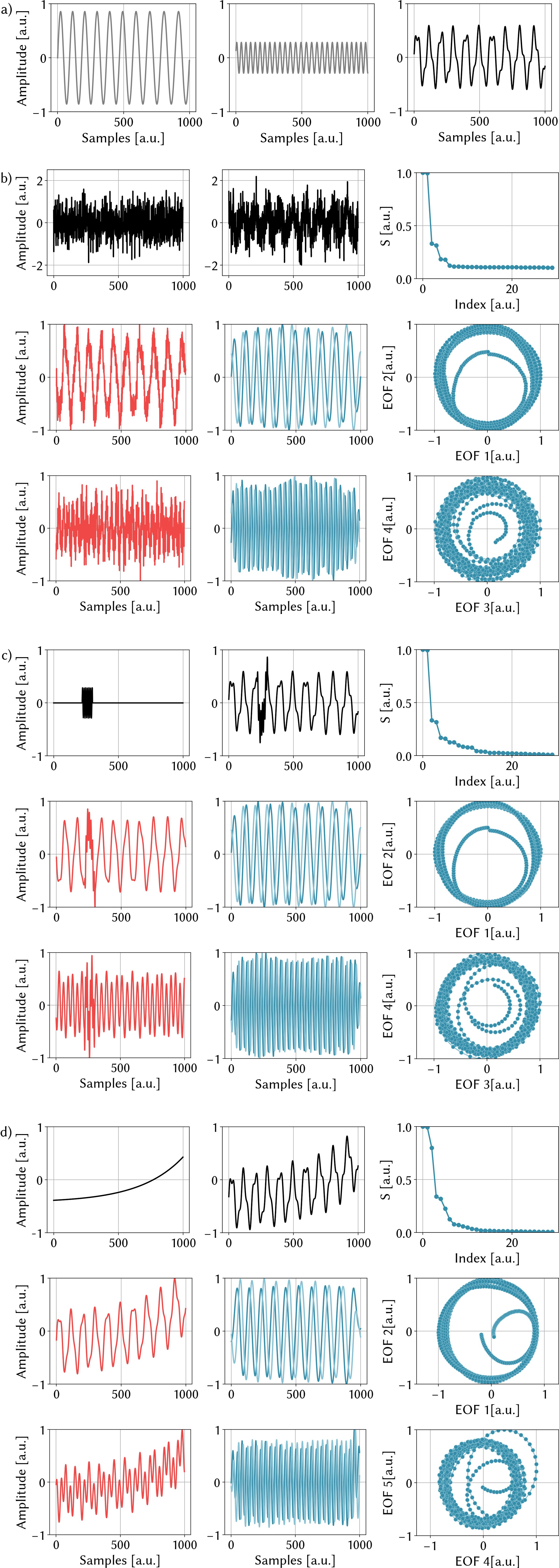}
	\caption{Comparison of PCA eigenvectors and SSA-FARY EOFs. a) 
	Frequency-modulated sinusoidal 
	oscillations ($a(t)$ left, $b(t)$ center) used 
	for the numerical simulations and weighted composite signal (right). 
	b), c), d) \textit{Black}: Different additional 
	signals (left) - b) noise, c) oscillatory spell, d) exponential trend - 
	added to 
	the composite signal of a) and a selected channel of the resulting signal 
	$X_i(t)$ (right). \textit{Red}: 
	First two eigenvectors of the PCA. \textit{Blue}: SSA-FARY singular values 
	$S$ (scree plot) and EOF pairs 
	(amplitude-time plot 
	and amplitude-amplitude scatter plot). All plots are normalized.}
	\label{Fig:Exp1}
\end{figure}
Fig.\ \ref{Fig:Exp1} depicts the results of the numerical simulations.
PCA is able to extract, at least essentially, the shape of the two oscillations 
$a(t)$ and $b(t)$
from $\bm{X}^\text{noise}$ and $\bm{X}^\text{spell}$. However, both the noise 
and the oscillation spell are still evident in the resulting eigenvectors and 
corrupt the signal. PCA fails to produce a useful result for $\bm{X}^\text{trend}$.
While oscillation $a(t)$ is present in two principle eigenvectors, oscillation $b(t)$
is not distinctly separated in any of the eigenvectors.

In contrast, SSA-FARY extracts the oscillation signals with almost no spoiling 
residuals in all three investigated cases. Only at the borders deviations from 
the ideal signal can be observed. Notably, SSA-FARY does not only 
yield a one-dimensional signal of the temporal evolution of an oscillation, but 
preserves the phase information by
quadrature pairs, as can be appreciated from the amplitude-amplitude plots. 
The two EOFs corresponding to the same pair have very similar singular values.
In the scree plot of 
Fig.\ \ref{Fig:Exp1} b) and c), the first plateau corresponds to EOF 1 and 2, 
and the second plateau belongs to EOF 3 and 4. In Fig.\ \ref{Fig:Exp1} d), the 
first two plateaus correspond to EOF 1 and 2, and to EOF 4 and 5. Hence, 
SSA-FARY does not mix the trend into the oscillations but creates an additional 
EOF to account for it.\footnote{In
Supplementary Material Fig.\ 1-3 we provide results for simulations with 
different noise variance and trend and spell amplitudes, as well as a different 
frequency 
variation.  More information on these figures is provided in chapter I of the 
supplementary document.}

\FloatBarrier
\paragraph*{Single-slice imaging}

\begin{figure}[h]
	\centering
	\includegraphics[width=0.75\textwidth]{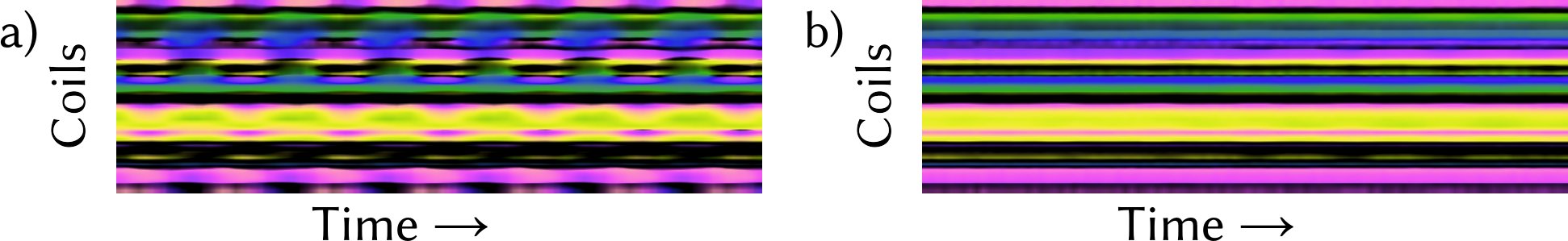}
	\caption{Snippet of the amplitude plot with color-coded phase of the DC 
	samples used for auto-calibration 
	of the single-slice reconstruction before (a) and after (b) 
	the data correction using the orthogonal projection. The period length 
	of the oscillations in (a) corresponds to 15 samples.}
	\label{Fig:AC}
\end{figure}

For each coil, Fig.\ \ref{Fig:AC} depicts the DC component of 100 consecutive 
spokes before and after the data correction 
using the orthogonal projection. Before correction some coils exhibit pronounced 
oscillations with approximately 15 samples per period. As we have used the seventh tiny 
golden  angle ($\varphi_0\approx 23.6^\circ$) the 
oscillations period of 15 samples corresponds to $\varphi^{15} = 15 \cdot 
\varphi_0 \approx 354^\circ$ (see eq.\ (\ref{Eq:Phi})). Hence, the 
oscillation period in the AC data is linked to the period of the projection 
angle. By removing this frequency and the higher-order harmonics 
these oscillations can be completely eliminated.

\begin{figure}[h]
	\centering
	\includegraphics[width=0.65\textwidth]{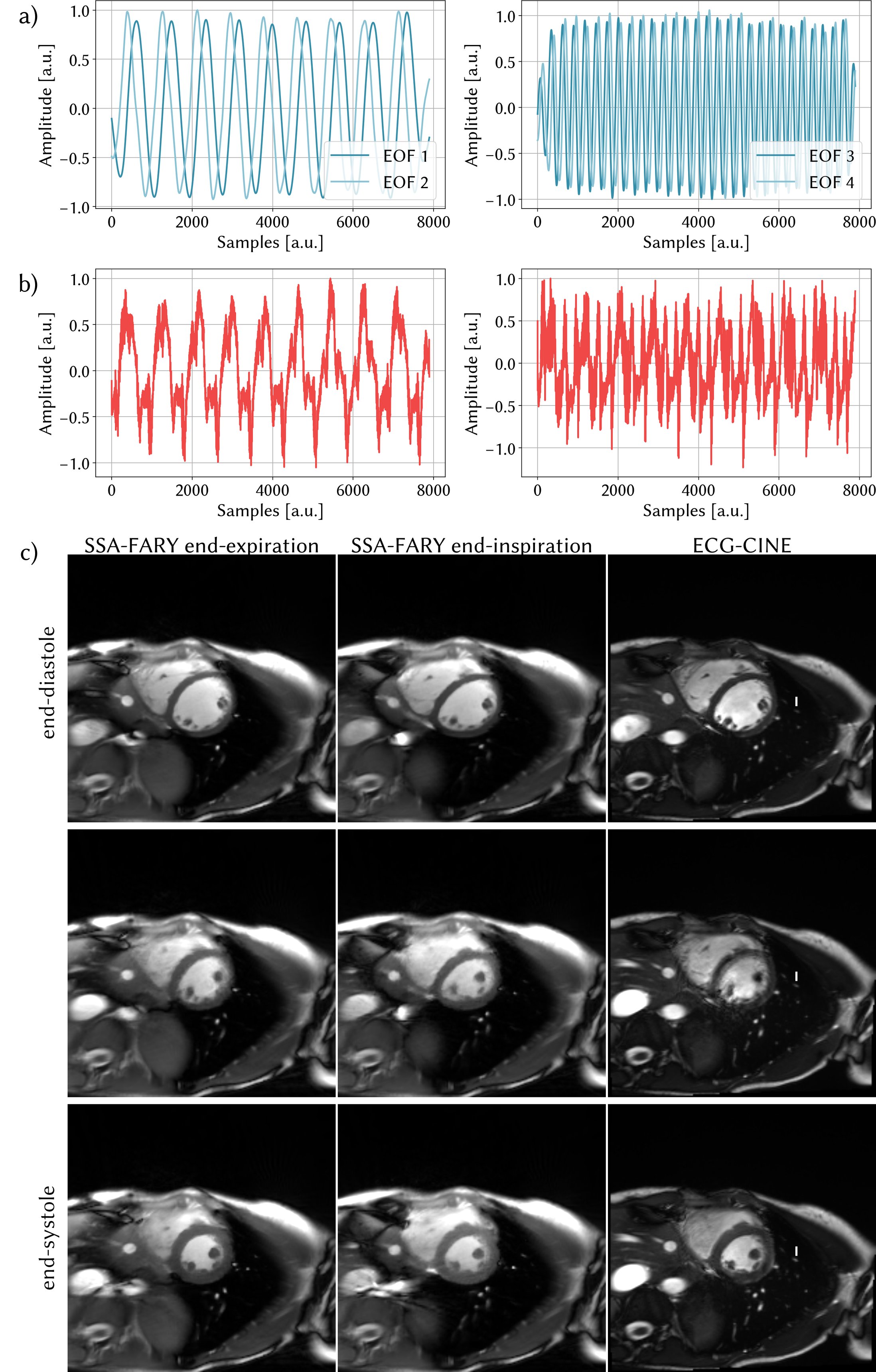}
	\caption{Self-gating signals and PICS reconstruction 
	of the human heart from a single-slice radial bSSFP acquisition
	in comparison to a conventional CINE scan.
	In (a) the two quadrature 
	EOFs of SSA-FARY are 
	plotted against time for 
	the respiratory (left) and cardiac (right) motion, respectively. In (b) the 
	first two eigenfunctions of the PCA are depicted. In both, a superposition 
	of the 
	cardiac and respiratory motion spoiled by additional noise can be perceived. For (c) the self-gating 
	signal of (a) was utilized. It shows 6 representative frames of the PICS 
	reconstruction
	using SSA-FARY
	corresponding to three cardiac phases from end-systole to end-diastole at end-expiration and 
	end-inspiration. For comparison, we also show the vendor images of an 
	ECG-triggered CINE 
	breath-hold scan.}
	\label{Fig:SS}
\end{figure}

Fig.\ \ref{Fig:SS} shows the self-gating signals generated with PCA
and SSA-FARY. In SSA-FARY, the first two EOFs represent 
respiratory motion and the third and fourth EOF 
cardiac motion. 
The EOFs of the pairs are in quadrature, respectively. Both the cardiac and the 
respiratory phases are well separated. In contrast, PCA cannot fully extract 
and separate the signals as respiratory and cardiac motion are superposed and 
heavily
spoiled by noise.

Fig.\ \ref{Fig:SS}c shows six representative images of the SSA-FARY-gated 
reconstruction. Depicted are from bottom to top the end-systolic, an intermittent and the end-diastolic frame for end-expiration and end-inspiration, 
respectively. For comparison, we also show the result of the ECG-triggered CINE 
breath-hold scan.

In Supplementary Material chapter II we present the results of a 
conventional 
gridding reconstruction of the full \deleted{90}72 second measurement gated 
with 
SSA-FARY.\footnote{Supplementary Material Fig.\ 4 shows the SSA-FARY and 
PCA 
gating signals of the 72 second scan. Moreover, a gridding reconstruction for 
three different cardiac phases is depicted and the results of a breath-hold and 
an
ECG-gated CINE scan are shown for comparison.} 

In Supplementary Material chapter III we present a similar experiment with a 
RF-spoiled gradient-echo sequence.\footnote{Supplementary Material 
Fig.\ 5 shows 
the effect of the proposed correction on the AC region. Supplementary Material 
Fig.\ 6 shows the self-gating signals determined with SSA-FARY and PCA. 
Furthermore, six representative frames of the SSA-FARY-gated reconstruction for 
different respiratory and cardiac motion states are depicted.}


\FloatBarrier
\paragraph*{SMS imaging}

\begin{figure}[h]
	\centering
	\includegraphics[width=0.85\textwidth]{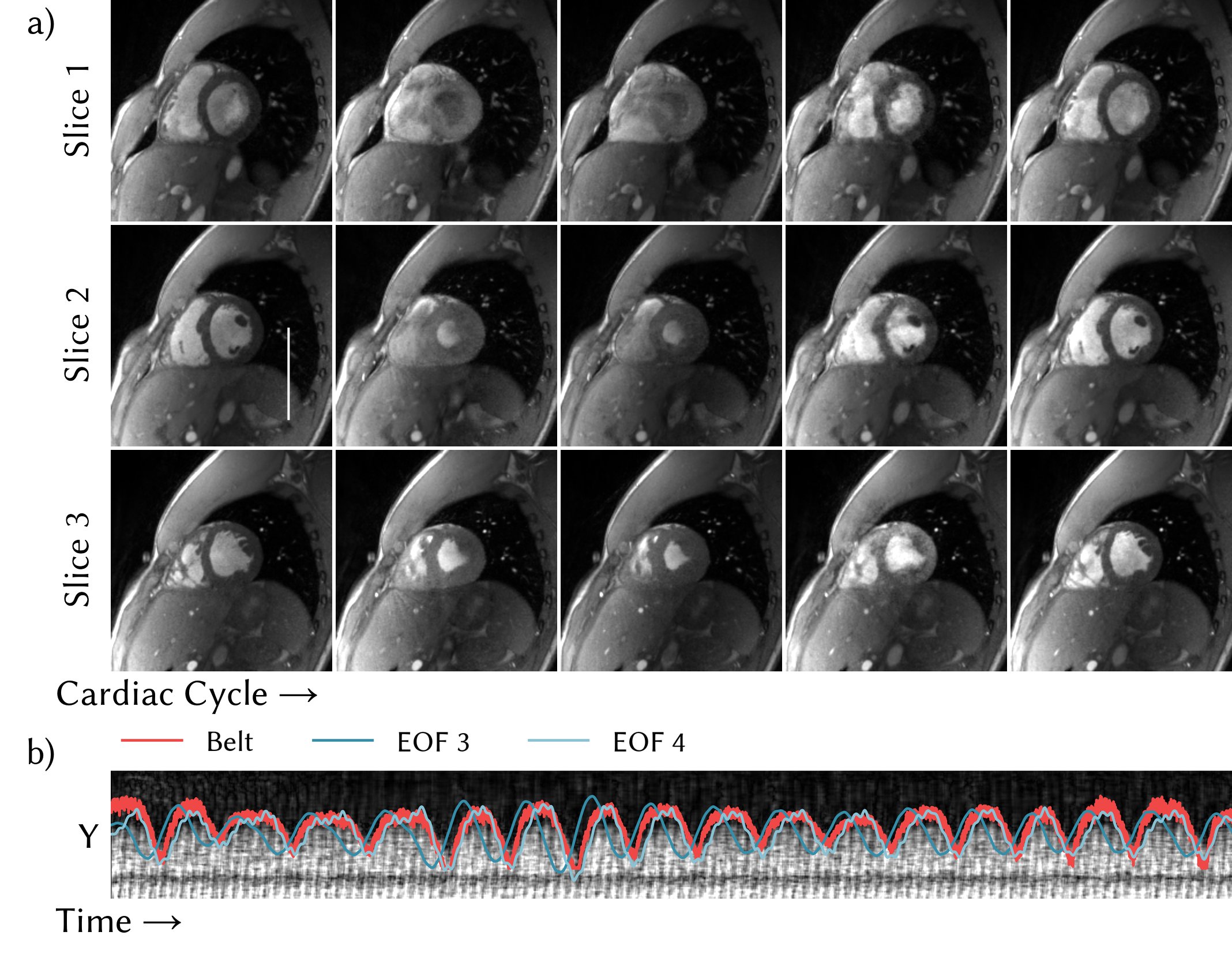}
	\caption{Images of the human heart
	reconstructed using PICS and the SSA-FARY self-gating signal
	from a radial SMS gradient-echo acquisition. (a) shows 5 out of 25 
	cardiac phases in end-expiration for each of the three slices. 
	The background of (b) is extracted from an SMS-NLINV 
	real-time
	reconstruction of 
	the entire time series and shows the temporal evolution of a vertical 
	line (highlighted in white), placed on the diaphragm in slice two. On 
	top, the respiratory 
	self-gating EOF quadrature pair of
	SSA-FARY and the signal extracted from the respiratory belt is 
	plotted. The amplitudes of the motion signals are scaled for the purpose 
	of comparison.}
	\label{Fig:SMS}
\end{figure}

Fig.\ \ref{Fig:SMS}a shows 5 out of 25 cardiac phases for all 
three 
slices in end-expiration. The different systolic and diastolic phases are well 
resolved.

The background of Fig.\ \ref{Fig:SMS}b shows the temporal evolution of a line 
extracted from an SMS-NLINV real-time
reconstruction of the full time 
series. The line was placed in slice two in vertical 
direction such that the actual motion of the diaphragm can 
be observed. This diaphragmatic motion is linearly related to the 
translation of 
the heart during breathing \cite{Wang_Magn.Reson.Med._1995}. On top of the 
background image, 
we have plotted the respiratory EOF quadrature pair obtained from SSA-FARY 
as well as the signal provided by the respiratory belt. For the purpose of 
comparison, the motion signals were scaled according to the amplitude of the 
diaphragm motion in the background image.

One of the EOFs (light blue) coincides very well both in amplitude and 
phase with the 
temporal evolution 
of the diaphragm. This is in line with the filtering interpretation of 
SSA-FARY, which states that one EOF of the quadrature pair is in phase with the 
related 
underlying motion, whereas the other EOF (dark blue) constitutes the quadrature 
signal. The in-phase EOF is furthermore in good agreement with the motion 
signal provided by the respiratory belt.

The presented volunteer shows a largely periodic breathing pattern. In 
Supplementary Material chapter IV we present the results of the same experiment 
on another volunteer, which happend to exhibit a highly irregular respiration 
pattern during measurement.\footnote{ 
	In Supplementary Material Fig.\ 7 we provide results of the same experiment 
	conducted on a different volunteer. We depict 5 cardiac phases for all 
	three slices, as well as a comparison between the SSA-FARY respiratory 
	self-gating
	signal, the respiratory belt and the diaphragm motion extracted from a 
	real-time reconstruction.}

\begin{figure}[h]
	\centering
	\includegraphics[width=0.48\textwidth]{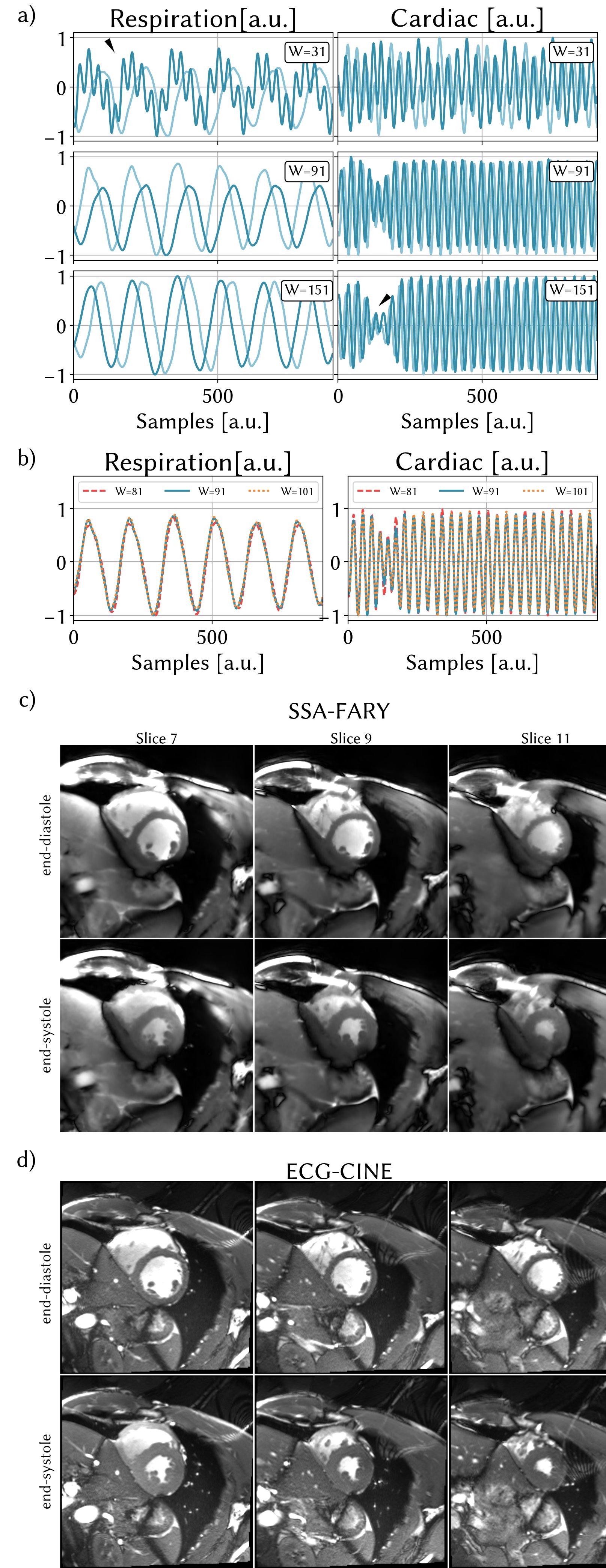}
	\caption{SSA-FARY self-gating signals for different window sizes
	from a SOS bSSFP acquisition and 
	representative slices from a 3D PICS reconstructions for volunteer V5
	in comparison to a conventional CINE scan.
	(a) A zoomed view on normalized EOF pairs  representing 
	respiratory and cardiac motion are depicted for window sizes $W=31$, $W=91$ 
	and $W=151$. The black arrows indicate signal mixing (top-left) and signal 
	loss (bottom right). (b) For window sizes $W=81$, $W=91$ and $W=101$, one 
	EOF 
	representing respiratory motion and one EOF representing cardiac motion are 
	plotted in a zoomed view on top of each other. (c) shows three 
	out of fourteen slices of an end-diastolic and end-systolic human heart 
	after expiration. (d) depicts the corresponding vendor images for the 
	ECG-triggered CINE breath-hold measurement.}
	\label{Fig:SOS}
\end{figure}

\FloatBarrier
\paragraph*{SOS imaging}
Fig.\ \ref{Fig:SOS}a depicts a zoomed view on EOFs representing 
respiratory 
and 
cardiac motion for different window sizes. For window size $W=31$, which - 
considering the undersampling scheme - covers a period of $\approx 
\SI{1.0}{\second}$, respiratory and cardiac motion are not fully separated and 
appear superposed in one EOF for respiratory and one EOF for cardiac motion. In 
contrast, for the proposed window size $W=91$ ($\approx \SI{2.8}{\second}$) the 
motion signals are well separated. Then again, for $W=151$ ($\approx 
\SI{4.5}{\second}$) a signal loss can be observed in the cardiac EOFs.
Fig.\ \ref{Fig:SOS}b shows one respiratory EOF and one cardiac EOF for the 
windows $W=81$, $W=91$ and $W=101$, covering periods from $\SI{2.5}{\second}$ 
to $\SI{3.0}{\second}$. All signals are in good agreement. The corresponding 
pairing-components of the EOFs show similar behavior and are therefore not 
depicted. 
Fig.\ \ref{Fig:SOS}c presents end-diastolic and end-systolic frames for three 
out of fourteen slices in end-expiration. All 
respiratory and cardiac states are well separated, the cardiac wall and the 
diaphragm are sharply resolved. Note, however, that some slices at the fringe 
of the slab have low signal intensity due to an unoptimized excitation profile. 
The image quality is comparable to the CINE reconstruction Fig.\ 
\ref{Fig:SOS}d, although the latter 
tends to be sharper. Due to the higher flip-angle, which is restricted by SAR
limitations in volumetric sequences, the CINE images possess a better blood 
myocardium contrast.

For the different volunteers, Table (\ref{Tab:ECG}) shows the average heart-rate 
$\bar{f}_\text{heart}$, 
its standard deviation $\sigma_\text{heart}$, the standard deviation 
$\sigma_\text{trig}$ of the synthetic SSA-FARY trigger to the ECG trigger and 
the end-diastolic and end-systolic left-ventricular blood-pool area of a 
mid-ventricular slice for CINE and 
SSA-FARY.

For imaging on a 3T system an insufficient shim can lead to banding 
artifacts. Measurements with bandings affecting the heart could not immediately 
be noticed and repeated as the reconstruction was performed offline. Therefore, 
these 
measurements were discarded for which the
analysis of one volunteer is omitted.

The observed average heart rates range from $\SI{0.91}{\hertz}$ 
($\SI{54.60}{}\,\text{bpm}$) to 
$\SI{1.38}{\hertz}$ ($\SI{82.20}{}\,\text{bpm}$) with different heart rate 
variabilities. For the given SOS 
acquisition with 14 partitions (6 AC lines and undersampling factor of 4), the 
temporal resolution of the SSA-FARY trigger is $\SI{30.4}{\milli\second}$. The 
standard deviation of the SSA-FARY trigger to 
the ECG trigger, $\sigma_\text{trig}$, is of similar size. Hence, 
the SSA-FARY trigger is in good agreement with the ECG signal and also matches 
the temporal resolution of the ECG-CINE acquisition.

The areas of the chosen mid-ventricular slices are comparable for ECG-CINE 
and 
SSA-FARY, particularly for end-diastole the difference lies mostly in the lower 
single-digit percent range, whereas a larger uncertainty can be observed for 
end-systole.

Volunteer V6 exhibits a highly erratic breathing pattern and volunteer V7 
possesses a strongly irregular heartbeat and furthermore unintentionally 
yawned three times during the measurement. Still, SSA-FARY can provide 
satisfying results as Table \ref{Tab:ECG} and the figures in the Supplementary 
Material show.\footnote{Supplementary Material Fig.\ 8 and 9 display the 
SSA-FARY gating signal, representative frames of the SSA-FARY-based image 
reconstruction and the corresponding CINE reconstructions for volunteers V6 and 
V7.  More information on the figures is provided in chapter V of the 
supplementary document.}

\begin{table*}[h]
	\centering
	\caption{Average heart rate  $\bar{f}_\text{heart}$ and corresponding 
		standard deviation $\sigma_\text{heart}$, standard deviation of SSA-FARY 
		from ECG trigger 
		$\sigma_\text{trig}$. End-systolic and end-diastolic left-ventricular 
		blood-pool area of 
		a mid-ventricular slice for ECG-CINE and SSA-FARY and corresponding 
		error. $^*$ 
		denotes the volunteer with highly erratic respiration, $^{**}$ denotes the 
		volunteer who yawned.
	}
	\resizebox{\textwidth}{!}{
	\begin{tabularx}{1.5\textwidth}{c ccc cccccc}
		\toprule
		\textbf{Volunteer} &  $\bm{\bar{f}_\text{heart}}$ [$\SI{}{\hertz}$] & 
		$\bm{\sigma_\text{heart}}$ [$\SI{}{\hertz}$]	& $\bm{\sigma_\text{trig}}$ 
		[$\SI{}{\milli\second}$] & \multicolumn{3}{c}{\textbf{Diastole} 
			[$\SI{}{\milli\meter\squared}$]}  
		& \multicolumn{3}{c}{\textbf{Systole} [$\SI{}{\milli\meter\squared}$]}\\
		\cmidrule(lr){5-7} \cmidrule(l){8-10} 
		&&&& ECG & SSA-FARY & Err. [$\%$] & ECG & SSA-FARY & Err. [$\%$] \\
		\cmidrule(lr){1-10} \\
		V1 & 0.910(4) & 0.057(3) & 23(1) & 2254 & 2222 & 1.4 & 1068 & 1017& 4.8 
		\\ 
		%%MRT5_MyT1_0009/00367  | 2c
		V2 & 1.380(3) & 0.0541(2) & 14.7(7) & 1536 & 1647& 
		7.2 & 573 & 706&  
		23.2      \\ 	
		%%MRT5_MyT1_0010/00590  | 3b    
		V3 & 0.979(2) & 0.0245(1) & 28(2) & 1911 & 2161 & 
		13.1 & 1072 & 1168 & 
		11.0\\ 
		%%MRT5_FR4D_0019/00245 | 4
		V4 & 1.245(3) & 0.045(2) & 23(1) & 1998 & 1956 &2.1& 866 & 961 & 11.0\\ 
		%%MRT5_FR4D_0020/00050 | 5b
		V5 & 0.955(4) & 0.056(3) & 19(1) & 2348 & 2380 & 1.4 & 1005 & 1179 & 
		17.3	\\ 
		%%MRT5_FR4D_0023/00314 | 8b
		V6$^*$ & 1.195(7) & 0.103(5) & 34(2) & 2244& 2220 & 1.1 & 1351 & 1407 & 
		4.1 \\ 
		%%MRT5_FR4D_0022/00038 | 7 
		V7$^{**}$ & 1.08(1)& 
		0.129(7)& 
		27(1) & 3398& 3314 & 2.5 & 2060  & 
		1930 & 6.3  \\ 
		%%MRT5_FR4D_0018/00427 | acc
		\bottomrule
	\end{tabularx}}
	\label{Tab:ECG}
\end{table*}

For all in vivo experiments, we have attached representative movies as 
Supplementary Material.\footnote{The files Mov1-14 of the Supplementary 
Material 
show representative movies of all in vivo reconstructions. More information on 
the movies is provided in chapter VI of the supplementary document.}

\FloatBarrier
\section{Discussion}

We introduced a novel dimensionality reduction method dubbed "SSA-FARY", 
which is based on Singular Spectrum Analysis and  showed that the proposed
technique can successfully recover the 
cardiac and respiratory signal from the AC data of single-slice, SMS and SOS 
MRI 
measurements.  Moreover, we have proposed an extended
orthogonal projection to correct for system imperfections in the AC data.

\paragraph*{AC correction}
The reasons for the oscillations in the AC data are many fold and 
according to 
our experience cannot be eliminated completely by techniques that correct for 
trajectory errors only \cite{Seiberlich_Magn.Reson.Med._2007, 
Deshmane_Magn.Reson.Med._2016}. Particularly for bSSFP sequences, eddy current related dephasing of spins additionally compromises the data \cite{Bieri_MagnResonMed_2005, Bieri_Magn.Reson.Med._2005}, although this effect should be rather small 
for the tiny golden angle \cite{Wundrak_IEEETransMedImag_2015}.

Zhang et al. \cite{Zhang_Magn.Reson.Med._2018} find this to be particularly 
problematic for 3D bSSFP imaging on a 3T system and propose to 
eliminate these measurement errors  by 
averaging  over the central five samples of each spoke. However, we
found this to produce even more oscillations in the AC data, which requires 
additional filtering.
In contrast, the orthogonal projection used here can remove most of the signal 
perturbations directly. For constant increments of the projection
angle, the correction can be thought of as a set of sharp 
band-stop or notch filters corresponding to higher-order harmonics
of a base frequency, which can be calculated from the increment of
the projection angle.

Strictly speaking, the AC correction is not mandatory for SSA-FARY as
without filtering these spurious oscillations appear as 
additional EOFs. However, as these oscillations usually manifest distinctly in 
the AC region, the corresponding EOFs often possess the highest singular 
values. Consequently, they dominate the result of eq.\ (\ref{Eq:Minimization})  
which reduces the significance and accuracy of the components of interest, i.e. 
cardiac and 
respiratory motion, and complicates the analysis. This can be
easily avoided by using the proposed orthogonal projection, which corrects for 
first-order system imperfections.

Alternatively, advanced techniques to measure the trajectory error 
\cite{Duyn_J.Magn.Reson._1998, Barmet_Magn.Reson.Med._2008} or higher-order 
system imperfection corrections could be utilized to account for the oscillations in the AC region \cite{Vannesjo_Magn.Reson.Med._2013, 
Stich_Magn.Reson.Med._2018}. Still, these approaches require additional 
hardware and/or sequence modifications, which limits their accessibility.

\paragraph*{SSA-FARY}

The presented dimensionality reduction technique for time-series SSA-FARY can 
be considered a PCA applied to a time-delayed embedding of the AC 
data to exploit the locally low-rankness of dynamic time series. The 
Block-Hankel matrix $\bm{A}$ (eq.\ (\ref{Eq:Hankel})) consists of shifted 
segments of the original time series. Its covariance matrix  
$C_t^{\;t'} = A_t(A^H)^{t'}$ (eq.\ 
(\ref{Eq:Cov})) is an array of scalar products specifying the correlation of 
all 
pairs of multi-channel segments in the embedding space. The considerable 
redundancy in the correlations of the segments in the presence of (quasi) 
repetitive oscillations, e.g. cardiac and respiratory motion, causes $\bm{C}$ 
to have low-rank. 
By exploiting not only spatial but also these temporal correlations, SSA-FARY 
performs better in recognizing temporal patterns than classical PCA and allows 
the separation of trend, oscillations and noise from the signal. This was 
successfully demonstrated on numerical simulations of superposed and spoiled 
sinusoidal time series. In the 
actual in vivo 
measurements the
respiratory and cardiac motion could be detected and clearly separated for all  
investigated sequence types.

To yield comparable results, methods like classical PCA must be combined with 
various pre- and post-processing techniques such as coil-selection or 
coil-clustering \cite{Zhang_Magn.Reson.Med._2016}, (iterative) band-pass 
filtering \cite{Liu_Magn.Reson.Med._2010} and signal smoothing 
\cite{Feng_Magn.Reson.Med._2016a}, which - especially for small AC regions - 
may be unstable and demand further manual tuning. By contrast, in SSA-FARY 
these steps are implicitly integrated and therefore surplus to requirement.
Moreover, since SSA-FARY preserves phase information by producing
quadrature pairs it allows for direct binning, which also renders
the otherwise mandatory 
peak detection obsolete \cite{Larson_Magn.Reson.Med._2004}.

We demonstrated the high quality of the SSA-FARY gating signals by 
comparison 
with a real-time reconstruction, a respiratory 
belt and ECG triggers. 
Moreover, at free breathing 
and at a considerably lower acquisition time SSA-FARY achieved a reconstruction 
quality that comes close to the results of the ECG-triggered CINE breath-hold 
scans. The analyzed left-ventricular blood-pool area of SSA-FARY
reconstructions mostly corresponds well to the CINE results for end-diastole,
whereas larger uncertainties were found for end-systole.
Here, because the left-ventricular blood-pool area is significantly
decreased compared to end-diastolic states, any deviation will result
in larger relative errors. One reason might be that due to the
high acceleration factor the short end-systolic phase is not
perfectly resolved. In addition, some discrepancy between a breath-hold scan 
and a gated free-breathing scan is expected, since it cannot be guaranteed 
that the selected respiratory bins exactly matches the anatomical state of
the breath-hold scan.

As the aim of this manuscript was the introduction of the self-gating 
technique, we did not fully optimize the sequence and 
reconstruction parameters, particularly the number of cardiac and respiratory 
bins, the regularization values of the ADMM and the undersampling scheme of the 
SOS sequence. The parameter 
tuning and the setup of a clinically applicable protocol is left for future 
investigations.

In single-slice imaging 
we only use a single sample per time step for auto-calibration and despite
the  proportionally large window size, which reduces statistical 
significance, SSA-FARY yields reliable results. Still, SSA-FARY 
	tends to be more resilient when multiple partitions are used and thus more 
	AC 
	data is available,
	as in SMS or SOS experiments, or when more samples relative to the 
window size are used for auto-calibration. For SMS or SOS imaging a bSSFP 
sequence is recommended since for RF-spoiled gradient-echo imaging a loss of contrast in systolic 
phases can occur 
due to pre-saturated blood flowing in from other slices, see Fig.\ 
\ref{Fig:SMS}. Note, however, that bSSFP sequences suffer from SAR limitations due to the increased flip-angle of the
RF pulse and are prone to banding artifacts when no proper
shimming is conducted.

Due to the zero-padding operation, eq.\ 
(\ref{Eq:ZeroPad}), and 
the subsequent Hankelization, eq.\ (\ref{Eq:Hankel}), the first and last 
samples in the SSA-FARY EOFs suffer from slight approximation errors. Still, 
for all presented experiments and analysis we did not discard these samples.

Occasionally, the trend was not completely separated from the EOFs for 
cardiac and respiratory motion, which we fixed by standardly using a moving 
average filter. The reason for the incomplete separation of trend can be 
understood 
by considering eq.\ (\ref{Eq:BP}), which relates the window size $W$ to 
the frequency bandwidth $\delta f_\text{B}$ of the eigenfilters $\bm{V}$ 
generating 
the EOFs $\bm{U}$. Our default choice of window size and sampling rate 
corresponds to $\delta 
f_\text{B}\approx 
\SI{0.35}{\hertz}$ for all measurements and sequences. If we choose $W$ too 
small, 
the EOFs 
capture a wider range of frequencies, which can result in a mixing of trend and 
oscillations. Similarly, if the respiration frequency (usually 
$f_\text{resp}\approx \SI{0.3}{\hertz}$) and cardiac frequency (usually 
$f_\text{resp}\approx \SI{1.0}{\hertz}$) happen to be spectrally close to one 
another, a very small window and the corresponding large frequency pass-band of 
the filters can hinder a 
proper separation, as it is the case in Fig.\ (\ref{Fig:SOS}a), $W=31$). Then 
again, if we choose a very large
$W$, the frequency response $f_B$ of the filters can be too narrow. 
Thus, possible frequency variations in the cardiac and respiratory motion can 
no longer be captured by a single (band-limited) EOF, which results in 
signal voids and/or the generation of additional EOFs representing higher 
order harmonics. Detecting these related EOFs corresponds to the 'Grouping' 
step 
of conventional SSA \cite{Golyandina__2013}. In the cardiac signal of Fig.\ 
(\ref{Fig:SOS}a) such a signal void caused by a cardiac frequency variation of 
$\approx \SI{0.3}{\hertz}$ can be observed for $W=151$, whereas the signal can 
still be 
adequately captured with $W=91$. As a summary, we found  $\delta 
f_\text{B}\approx 
\SI{0.35}{\hertz}$ to be a robust choice and we propose to choose the window 
size accordingly using eq.\ (\ref{Eq:BP}).

The computational bottle-neck of SSA-FARY is the SVD of 
$\bm{C}$ of size $[N_t \times N_t]$.
Especially for single-slice imaging, the window size $W$ required
to obtain $\delta f_\text{B}\approx \SI{0.35}{\hertz}$ and 
the corresponding number of AC samples $N_t$ to obtain good results
is relatively large. Hence, the decomposition 
of $\bm{C}$ is rather time-consuming. Nevertheless, there are approaches to 
significantly speed up the decomposition stage 
\cite{Korobeynikov_StatisticsanditsInterface_2010, Leles_SoftwareX_2018}. 

SSA-FARY 
reliably detected the EOF pairs corresponding to 
cardiac and respiratory motion, which for RF-spoiled gradient echo measurements 
usually possess 
the 
highest singular values. It is, however, not determined that the two largest 
components belong to the cardiac or the respiratory motion. For bSSFP 
sequences, which are generally more prone to system imperfections, we 
frequently found other components such as trends to have high singular values, 
too. Although for this study the components used for gating were chosen by 
visual 
inspection of the EOFs, it is fairly simple to automatize the assignment 
using a frequency analysis.

If one finds the separation of SSA-FARY to work insufficiently, a variation 
of 
the 
window 
size usually helps to recover a suitable gating result. Still, for patients 
with strong 
cardiac or respiratory frequency variations, highly non-periodic respiratory 
motion or arrhythmia the results of the presented SSA-FARY method might still 
be 
insufficient. In this case various promising extensions 
of SSA(-FARY) exist to improve the results, e.g. 'Nonlinear 
Laplacian Spectral Analysis', 'Sliding SSA', 'Oblique SSA', 'Nested SSA' and 
their combinations \cite{Giannakis_PNAS_2012, 
Harmouche_IEEETrans.SignalProcessing_2018, Golyandina_Stat.Interface_2015}.

\paragraph*{Outlook} The property of SSA-FARY to not only reliably separate 
cardiac and respiratory 
motion but also to extract the trend in data, suggests further applications.  
In dynamic 
contrast-enhanced MRI, SSA-FARY could replace spline-fitting
\cite{Feng_Magn.Reson.Med._2016a} for separating motion and
contrast enhancement. Furthermore, preliminary results suggest that 
the T1 decay in inversion recovery sequences is detected as an individual 
component and separated 
from the motion signals, which would enable the simultaneous reconstruction of 
parameter maps and self-gated anatomical motion 
\cite{Christodoulou_NatureBiomedicalEngineering_2018}.

Last but not least we want to mention that the zero-padding approach used
in SSA-FARY is not limited to self-gated MRI but turns multi-variate SSA
into a general dimensionality reduction method, which might be useful
for many other problems related to time-series analysis.

\section{Conclusion}
We have introduced a novel SSA-based dimensionality reduction method called 
SSA-FARY. Its intuitive approach, the easy implementation and its capability to 
separate cardiac and respiratory motion as well as trend makes it a promising 
approach for MRI self-gating, particularly when it is combined with efficient 
data acquisition schemes and a state-of-the-art image reconstruction technique.

\section{Acknowledgement}
The orthogonality projection to correct the AC data was originally conceived 
by Dr. Anthony G. Christodoulou (Biomedical Imaging Research Institute, Cedars-Sinai 
Medical Center). We extended it by including the higher-order harmonics.

\nottoggle{IEEE}
{
	\Funding
}

\appendix
\section{Details for Numerical Simulations}
Here we present the values for the variables used in the numerical 
simulation (Fig.\ \ref{Fig:Exp1}). The total duration $T=800\; 
\text{[a.u.]}$ 
consists 
of $1000$ discrete samples. 
The other parameters can be found in Table \ref{Tab1}.

\begin{table}[h!]
	\centering
	\caption{Variables and values for the numerical simulation.}
	\begin{tabular}{cc}
		\toprule		
		\textbf{Variable} 	& \textbf{Value} 			\\
		\midrule
		$N_c$		& $30$				\\
		$A$ 		& $3$ 				\\
		$\phi_a$ 	& $0$ 				\\
		$\omega_a$ 	& $\frac{2\pi}{80}$ \\
		$B$ 		& $1$ 				\\
		$\phi_b$ 	& $0.5$ 			\\
		$\omega_b$ 	& $\frac{2\pi}{30}$ \\
		$\sigma_\text{noise}$ & $2$	\\
		$C$ 		& $1$ 				\\
		$\phi_c$ 	& $0$ 				\\
		$\omega_c$ 	& $\frac{2\pi}{10}$ \\
		$[t_1,t_2]$ & $[220,300]$		\\                    
		$D_1$		& $0.15$				\\
		$D_2$		& $1.5$				\\
		$\xi$		& $3.75\cdot 10^{-3}$				\\		
		$W$			& $101$ 			\\
		$\Phi$      & $2$ \\
		\bottomrule
	\end{tabular}
\label{Tab1}
\end{table}

%\newpage
\bibliographystyle{IEEEtran}
\bibliography{radiology}

\iftoggle{ARXIV}
{


\section*{Supplementary material}

# Supplementary Document - Cardiac and Respiratory Self-Gating in Radial MRI using an Adapted Singular Spectrum Analysis (SSA-FARY)

Sebastian Rosenzweig[*,1,2], Nick Scholand[1,2], H. Christian M. Holme[1,2], and Martin Uecker[1,2]

[1]Institute for Diagnostic and Interventional Radiology, University Medical Center Göttingen, Göttingen, Germany
[2]German Centre for Cardiovascular Research (DZHK), Partner Site Göttingen, Göttingen, Germany

March 9, 2020

## Abstract

This supplementary material provides additional information and experiments regarding the numerical simulation and the in vivo experiments.

## 1 Numerical Simulations with Varying Simulation Parameters

We demonstrate the effects of different parameter values on the numerical simulation. For each simulation, we keep the basic parameters as given in Table II of the main part of the manuscript and list all modified parameters in Table 1 of this supplementary material.

For Fig. (1), we moderately increase the noise variance and the amplitude of the spell and the trend. Consequently, the PCA performs even worse in detecting the individual oscillations, particularly for the oscillation with the higher frequency. In contrast, SSA-FARY still provides satisfying results.

---

[*]Sebastian Rosenzweig, University Medical Center Göttingen, Institute for Diagnostic and Interventional Radiology, Robert-Koch-Str. 40, 37075 Göttingen, Germany. sebastian.rosenzweig@med.uni-goettingen.de

1

Table 1: Variables and values for the numerical simulation.

|  | Variable | Value |
|---|---|---|
| Fig. (1) | $\sigma_{\text{noise}}$ | 4 |
|  | $C$ | 2 |
|  | $D_1$ | 0.25 |
|  | $D_2$ | 2.5 |
|  |  |  |
| Fig. (2) | $\sigma_{\text{noise}}$ | 7 |
|  | $C$ | 3 |
|  | $D_1$ | 0.35 |
|  | $D_2$ | 3.5 |
|  |  |  |
| Fig. (3) | $\Phi$ | 3 |

For Fig. (2), we further increase the noise variance and the amplitude of the spell and the trend. Here, the SSA-FARY result for the lower frequency oscillations are still acceptable, whereas the EOFs for the higher frequency oscillation are now also clearly corrupted.

For Fig. (3), the same parameters for noise variance, trend and spell are used as in the main part of the manuscript, but a higher frequency variation is simulated. The PCA result is comparable to the one in the main part of the manuscript, while the SSA-FARY results look still satisfying, yet slightly worse due to the limited frequency bandwidth of the EOFs, which is determined by the window size $W$.

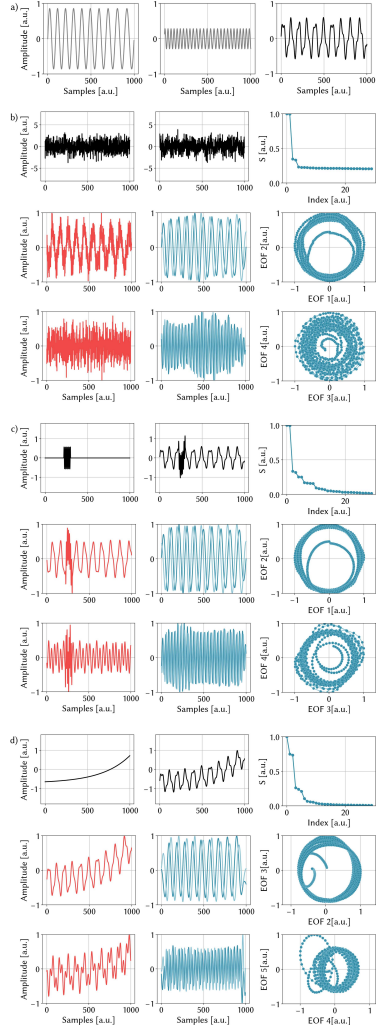

Figure 1: Comparison of PCA eigenvectors and SSA-FARY EOFs. In this example there is a moderate increase of the noise variance and the spell and trend amplitude compared to the main part of the manuscript. a) Frequency-modulated sinusoidal oscillations ($a(t)$ left, $b(t)$ center) used for the numerical simulations and weighted composite signal (right). b), c), d) *Black*: Different additional signals (left) - b) noise, c) oscillatory spell, d) exponential trend - added to the composite signal of a) and a selected channel of the resulting signal $X_i(t)$ (right). *Red*: First two eigenvectors of the PCA. *Blue*: SSA-FARY singular values $S$ (scree plot) and EOF pairs (amplitude-time plot and amplitude-amplitude scatter plot). All plots are normalized.

3

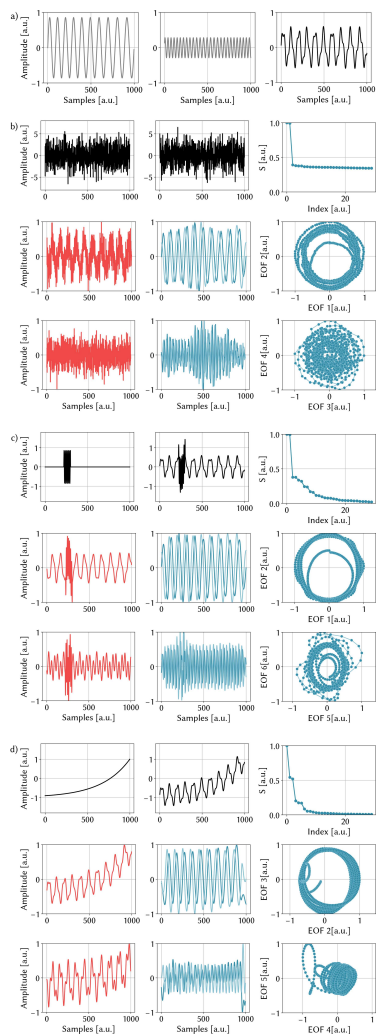

Figure 2: Comparison of PCA eigenvectors and SSA-FARY EOFs. In this example there is a strong increase of the noise variance and the spell and trend amplitude compared to the main part of the manuscript. See also caption of Fig. (1).

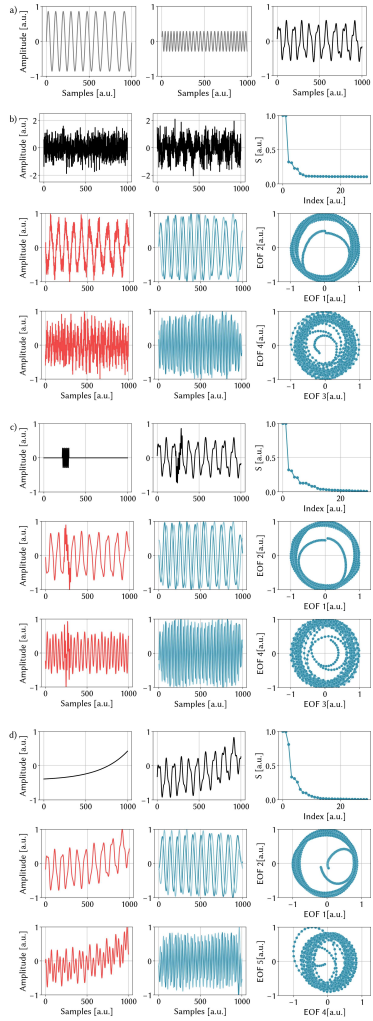

Figure 3: Comparison of PCA eigenvectors and SSA-FARY EOFs. In this example there is a moderate increase of the frequency variation compared to the main part of the manuscript. See also caption of Fig. (1).

5

## 2 SSA-FARY Gated bSSFP reconstruction using conventional gridding

While we have used only 30 seconds of data in the single-slice experiment described in the main part of the manuscript, we now apply SSA-FARY self-gating and PCA self-gating to the full 72 seconds of data. We use the same SSA-FARY window size $W = 751$ and bin the data into 25 cardiac and 3 respiratory bins. On the SSA-FARY-gated data we perform a conventional gridding reconstruction followed by a root-sum-of-square coil combination.

Similar to the results of the main part of the manuscript, SSA-FARY outperforms the PCA self-gating, which fails to separate the cardiac from respiratory motion, see Fig. 4.

Because of the longer acquisition time and the reduced number of respiratory bins, each cardiac state contains sufficient spokes such that a conventional gridding reconstruction can be performed. While mild undersampling artifacts can be observed, the temporal dynamics are well resolved, see Fig. 4 and movie Mov5.

Note, that due to the limited number of respiratory bins we can't depict the matching respiratory state of the ECG-gated CINE breath-hold scan for the gridding reconstruction.

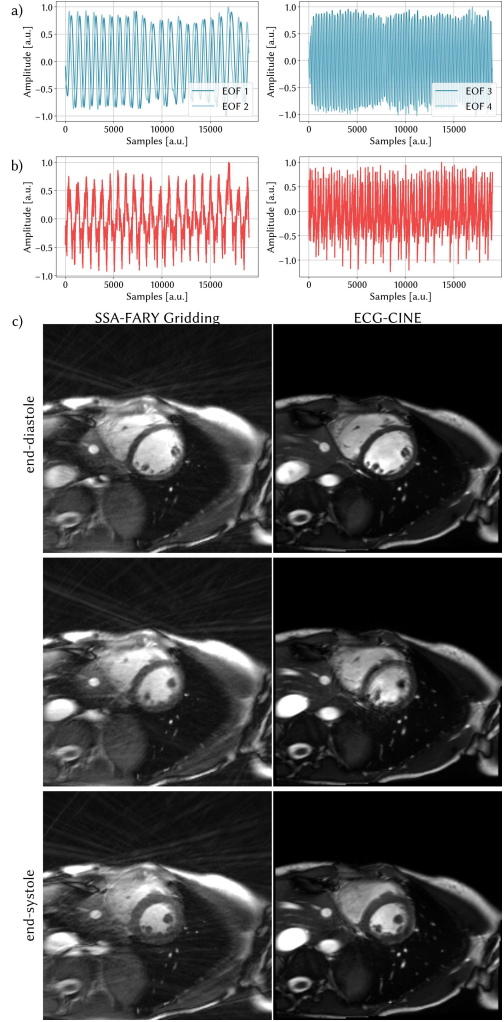

Figure 4: Self-gating signals and gridding reconstruction of a single-slice radial bSSFP aquisition in comparison to conventional CINE images. In (a) the two quadrature EOFs of SSA-FARY are plotted against time for the respiratory (left) and cardiac (right) motion, respectively. In (b) the first two eigenfunctions of the PCA are depicted. For (c) the self-gating signal of (a) was utilized. It shows three representative frames of the SSA-FARY-gated gridding reconstruction corresponding to three cardiac phases from end-systole to end-diastole for a single respiration state. As in the Figure 4 of the main manuscript, we also show vendor images of the ECG-triggered CINE breath-hold scan.

# 3 SSA-FARY Gated Reconstructions for a Single-Slice RF-Spoiled Gradient-Echo Sequence

Similar to the bSSFP single-slice experiment of the main part of the manuscript, we perform a 30 second RF-spoiled gradient-echo scan (TE/TR = 1.63/2.60 ms, flip angle 12°) with slice-thickness 7 mm of the human heart in short-axis view.

We compare the principal motion signals using PCA and SSA-FARY with window size $W = 1171$, which covers a period of about 3 s.

For each coil, Fig. 5 depicts the DC component of 100 consecutive spokes before and after the data correction using the orthogonal projection. Just as in Fig. 3 of the main part of the manuscript, some coils exhibit pronounced oscillations with approximately 15 samples per period, which is related to the seventh tiny golden angle. With the proposed correction method these oscillations can be eliminated.

Fig. 6 shows the self-gating signals generated with PCA and SSA-FARY. In SSA-FARY, the first two EOFs represent respiratory motion and the third and fourth EOF cardiac motion. The EOFs of the pairs are in quadrature and the motion signals are well separated. PCA fails in separating the cardiac from the respiratory motion and both principle components are spoiled by noise.

Fig. 6c shows six representative images of the SSA-FARY-gated reconstruction, three of which in end-systole and three in end-diastole. Depicted are three respiratory states from end-inspiration to end-expiration.

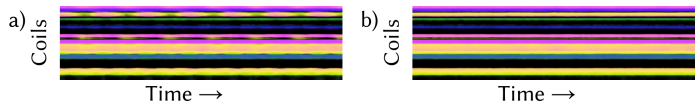

Figure 5: Snippet of the amplitude plot with color-coded phase of the DC samples used for auto-calibration of the single-slice reconstruction before (a) and after (b) the data correction using the orthogonal projection. The period length of the oscillations in (a) corresponds to 15 samples.

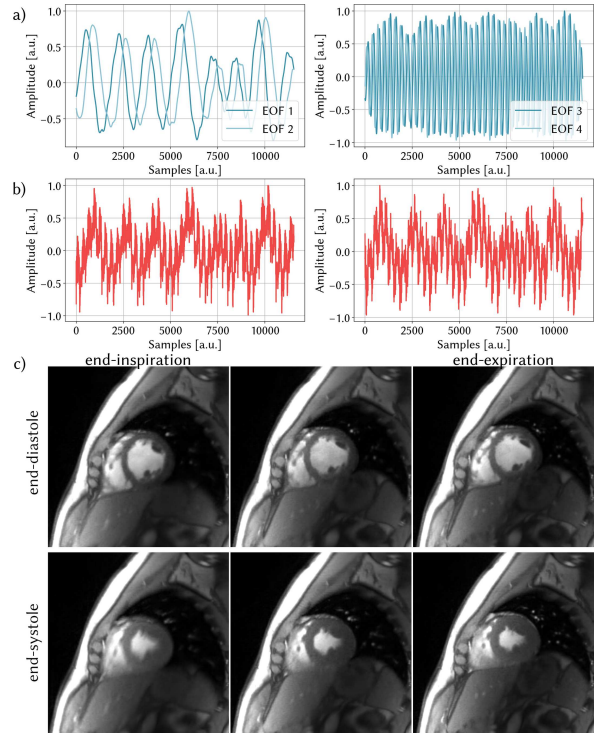

Figure 6: Self-gating signals and PICS reconstruction from a single-slice radial gradient-echo acquisition. In (a) the two quadrature EOFs of SSA-FARY are plotted against time for the respiratory (left) and cardiac (right) motion, respectively. In (b) the first two eigenfunctions of the PCA are depicted. In both, a superposition of the cardiac and respiratory motion can be perceived. For (c) the self-gating signal of (a) was utilized. It shows six representative frames corresponding to end-systole and end-diastole for end-inspiration, an intermittent state and end-expiration.

# 4 Respiratory Signal Comparison for Irregular Breathing

We perform the same SMS measurement as described in the main part of the manuscript on another volunteer who happened to have an irregular breathing pattern during the measurement. We depict the results of the experiment in Fig. 7.

For all three slices, Fig. 7a shows 5 out of 25 cardiac phases in end-expiration, where the individual cardiac phases can be distinguished well. From the SMS-NLINV real-time reconstruction of the full time series, we determine the actual motion of the lung by extracting a line from slice one which cuts the diaphragm in vertical direction, and use it as background for Fig. 7b. On top, we plot a respiratory EOF obtained using SSA-FARY. It shows good agreement with the diaphragm motion in the background.

We also depict the signal from the respiratory belt, which tends to be temporarily misaligned with the diaphragm position after deep inhalation.

For the purpose of clarity, we do not plot the quadrature pair belonging to the depicted EOF. Similar to Fig. 5 of the main part of the manuscript, the motion signals were scaled to match the amplitude of the diaphragm position displayed in the background image.

Deviations of the respiratory belt signal from the actual diaphragm position are not surprising. The correlation between external physiological motion and internal organ movement is subject dependent and patient specific signal processing can be required to account for the associated dephasing [1, 2], which is specifically important for irregular breathing patterns [3, 4]. Moreover, it is a know issue that the signal from a respiratory belt may be imperfect [5]. Furthermore, it was shown that an internal navigator based on the diaphragmatic motion outperforms the respiratory belt [6]. Consequently, the respiratory belt is not widely used for respiratory gating in MRI [7].

This experiment shows that - even for irregular breathing - the respiratory gating signal obtained with SSA-FARY resembles the diaphragmatic position largely well both in phase and in amplitude and therefore constitutes a suitable surrogate for respiratory gating.

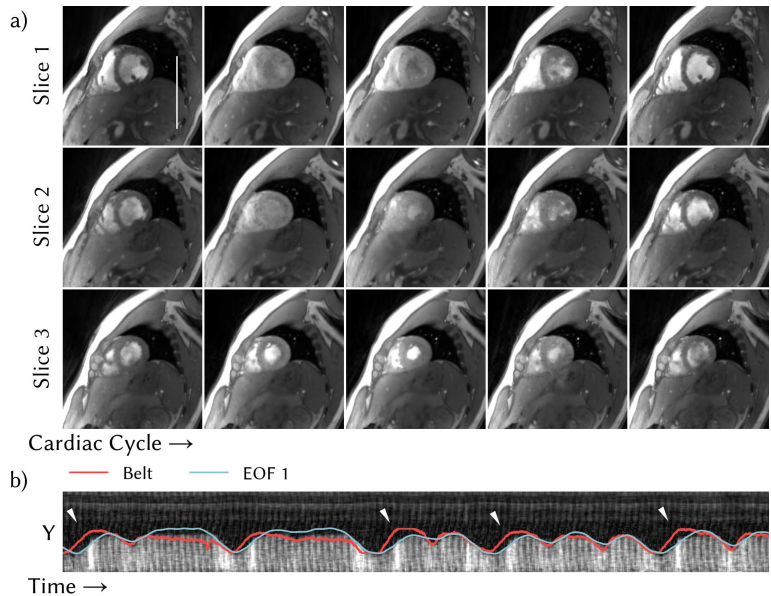

Figure 7: PICS reconstructions and self-gating signal of a radial SMS gradient-echo acquisition. For all three slices, (a) shows 5 out of 25 cardiac phases in end-expiration. The background of (b) is extracted from a SMS-NLINV real-time reconstruction of the entire time series. It shows the temporal evolution of a vertical line (highlighted in white), placed on the diaphragm in slice one. As an overlay, the signal obtained from the respiratory belt and the EOF which is in phase with the lung motion is plotted. The signal from the respiratory belt shows some disparities compared to the real-time reconstruction after deep inhalation. This is indicated by the white arrows.

11

# 5 SSA-FARY Gated Reconstructions for Highly Irregular Breathing, Yawning and a Pronounced Heart Rate Variability.

Here, we present more details on the gating and reconstruction of the Stack-of-Stars measurements of volunteers V6 and V7.

**Irregular breathing pattern** Volunteer V6 shows a non-dictated highly irregular breathing pattern with periods of normal breathing, interrupted by various breath-hold intervals of up to approximately 12 s. The default EOF bandwidth $\delta f_\mathrm{B} \approx 0.35\,\mathrm{Hz}$, which here corresponds to a window size of $W = 91$, is too narrow to capture this complex breathing motion in a single EOF pair. In fact, SSA-FARY creates at least 6 EOFs to account for the respiratory motion. Consequently, the cardiac motion is only given by EOF 8 and EOF 9, see Fig. (8a). Since for respiratory binning a single EOF pair is required, we run SSA-FARY again with reduced window size $W = 51$ to increase the spectral width of the EOFs. This choice and the corresponding EOFs 2 and 3 allow for a proper binning of the respiratory motion. The window size of the moving average was adjusted to $L_\mathrm{avg} = 600$ samples. The results for end-expiration are depicted in Fig. (8b), together with the corresponding slices of the ECG-CINE breath-hold measurement (8c).

**Yawning and pronounced cardiac frequency variations** Volunteer V7 admitted that he unintentionally yawned during the three minute measurement. The volunteer furthermore exhibits a highly irregular heartbeat with the heart rate standard deviation of $\bar{\sigma}_\mathrm{heart} = 0.129\,\mathrm{Hz}$ and heart rate jumps of up to $0.3\,\mathrm{Hz}$.

Using the window size $W = 91$, the three yawns are very well reflected in the first EOF, while the respiration is represented by EOFs 2 and 3, see Fig. (9a). The first EOF was used to discard the intervals in which the volunteer yawned before binning, which reduces the for the reconstruction effectively available data about 30 s. Note that the analysis for $\sigma_\mathrm{trig}$ of Table I in the main part of the manuscript was performed using the full data-set.

Due to the high heart rate variability, we run SSA-FARY again with window size $W = 21$ to increase the spectral width of the EOFs for cardiac binning. Although such a low window-size can lead to a mixing of respiratory and cardiac oscillations, as was shown in Fig. (6a) of the main part of the manuscript, this is not the case here and the determined gating signal is in good agreement with the ECG signal, see Table I of the main part of the manuscript. To better compensate the signal offset in the cardiac EOFs, which particularly arouse during the yawing periods, we adapted the moving average filter for the cardiac EOFs to $L_\mathrm{avg} = 30$.

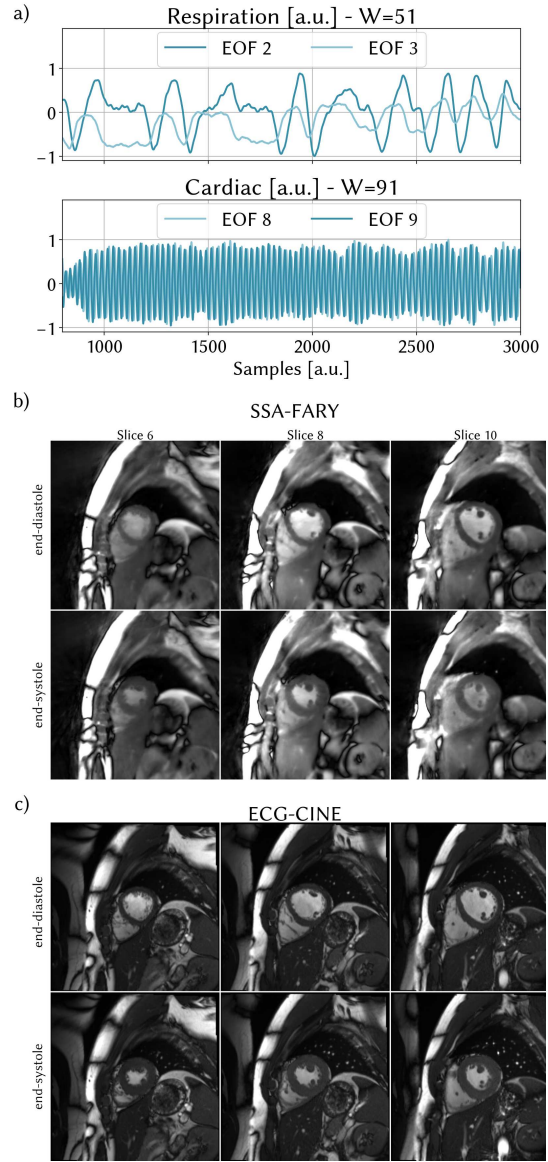

Figure 8: SSA-FARY self-gating signals and representative slices from a 3D PICS reconstruction of a SOS bSSFP acquisition for volunteer V6 in comparison to conventional CINE images. (a) depicts a zoomed view on normalized EOF pairs representing respiratory and cardiac motion using the window sizes $W = 51$ and $W = 91$, respectively. (b) shows three out of fourteen slices in end-diastole and end-systole after expiration. (c) shows the corresponding reconstructions for the ECG-triggered CINE breath-hold measurement.

13

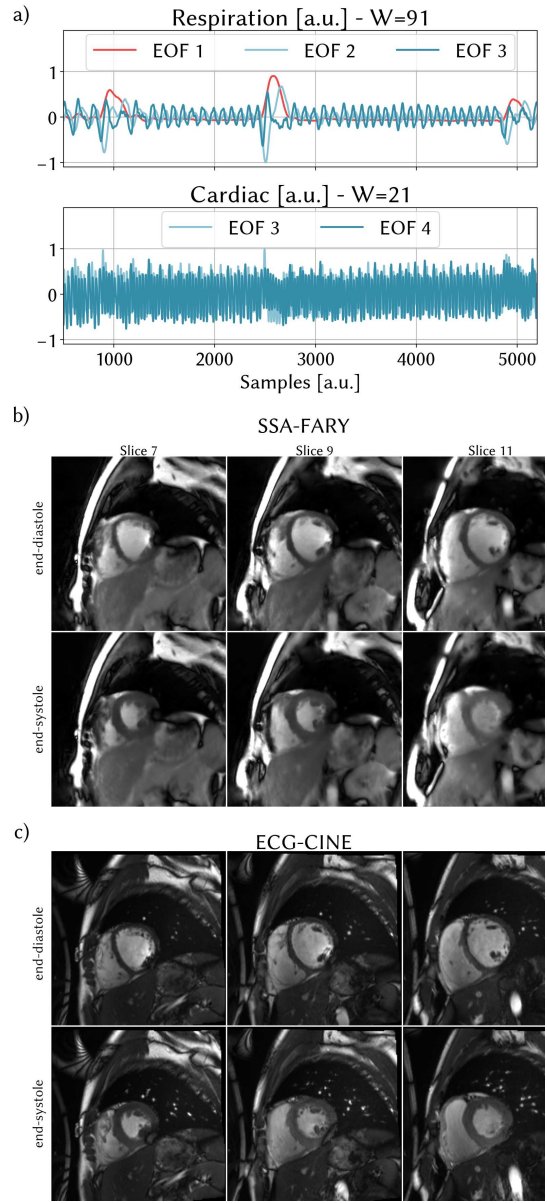

Figure 9: SSA-FARY self-gating signals and representative slices from a 3D PICS reconstruction of a SOS bSSFP acquisition for volunteer V7 in comparison to conventional CINE images. (a) A zoomed view on the normalized EOFs 1, 2 and 3 determined with window size $W = 91$. While EOF 1 captures the yawning, the EOFs 2 and 3 represent breathing motion. For window size $W = 21$, the EOFs 3 and 4 capture the cardiac motion. (b) shows three out of fourteen slices in end-diastole and end-systole after expiration. (c) shows the corresponding reconstructions for the ECG-triggered CINE breath-hold measurement.

# 6   Movies of the In Vivo Experiments

For the single-slice bSSFP (Mov1-2) and RF-spoiled gradient-echo measurements (Mov3-4), we have attached two movie files, respectively. Mov1 and Mov3 show the cardiac cycle for end-expiration (left) and end-inspiration. Mov2 and Mov4 show the respiratory cycle for end-systole (left) and end-diastole. To account for the low number of respiratory bins, we have interpolated the frames in the respiratory resolved videos.

Mov5 shows a cardiac cycle of the gridding reconstruction.

Mov6 shows a cardiac cycle for all three slices of the SMS reconstruction from the main part of the manuscript at end-expiration (top) and end-inspiration.

Mov7 shows a cardiac cycle for all three slices of the SMS reconstruction with erratic breathing at end-expiration.

For all analyzed volunteers V1-V7 of the Stack-of-Stars measurements, we have attached a representative movie (Mov8 - Mov14) showing a cardiac cycle for 6 out of 14 slices in end-expiration.


}

\end{document}